\documentclass[a4paper,11pt]{article}
\pdfoutput=1

\usepackage{jcappub}

\usepackage[T1]{fontenc}
\usepackage{graphicx,amssymb}
\usepackage{bbm}
\usepackage[utf8]{inputenc}
\usepackage{amsmath}
\usepackage{newunicodechar}
\usepackage{booktabs}
\DeclareMathOperator{\sech}{sech}

\title{\boldmath Dynamics of Biased Domain Walls: The Rocket Effect}

\author[a,b,1]{R. B. Vilhena\note{Corresponding author.}}

\author[a,b]{P.P. Avelino}

\author[a,c]{C. dos Santos}

\affiliation[a]{Departamento de F\'{\i}sica e Astronomia, Faculdade de Ci\^encias, Universidade do Porto, Rua do Campo Alegre 687, PT4169-007 Porto, Portugal}
\affiliation[b]{Instituto de Astrof\'{\i}sica e Ci\^encias do Espa{\c c}o, Universidade do Porto, CAUP, Rua das Estrelas, PT4150-762 Porto, Portugal}
\affiliation[c]{Centro de F\'{\i}sica das Universidades do Minho e do Porto, Faculdade de Ci\^encias da Universidade do Porto, 4169-007, Porto, Portugal}

\emailAdd{ricborvil@gmail.com, pedro.avelino@astro.up.pt, cssilva@fc.up.pt}

\abstract{We investigate the dynamics of domain walls in scalar field theories with degenerate vacua (i.e., vacua of equal energy density) in which the scalar field mass depends on the vacuum state. Using analytical arguments and numerical simulations, we show that this vacuum dependence of the scalar field mass renders the emission of scalar radiation from domain walls anisotropic, preferentially toward regions with smaller scalar field mass. We further show that the resulting recoil (rocket) effect biases the evolution of cosmological domain wall networks in favor of the lower-mass vacuum, thereby promoting network decay. We also demonstrate that the biased evolution of domain walls in theories with degenerate vacua, previously attributed to asymmetries of the potential barrier near the local maximum, is instead primarily controlled by the vacuum dependence of the scalar field mass. More generally, in theories with non-degenerate vacua, this recoil mechanism constitutes an additional source of dynamical bias that can either hasten or delay network decay relative to the standard expectation based solely on differences in vacuum energy density.}

\begin{document}
	\maketitle
	\flushbottom
	
\section{Introduction}\label{sec:intr}
    
Domain walls, which separate regions that settle into different vacuum states, are a generic consequence of the spontaneous breaking of discrete symmetries in quantum field theory. They are expected to form during cosmological phase transitions in the early Universe \cite{Zeldovich:1974uw,Kibble:1976sj,Vilenkin:2000jqa} and have long been studied because of their nonlinear dynamics and potential observational consequences. These include anisotropies in the cosmic microwave background \cite{Sousa:2015cqa,Caloni:2026dyu}, a stochastic gravitational-wave background \cite{Hiramatsu:2013qaa,NANOGrav:2023hvm,Gouttenoire:2023ftk,Ferreira:2024eru,Gruber:2024pqh,Notari:2025kqq,Babichev:2025stm,Cyr:2025nzf}, the generation of non-linear density perturbations that may seed the early formation of massive structures \cite{Winckler:2025hbc}, the production of compact objects such as primordial black holes \cite{Ferreira:2024eru,Dunsky:2024zdo}, and modifications to the expansion history of the Universe \cite{Bucher:1998mh,PinaAvelino:2006ia}. 

Owing to their $(2+1)$-dimensional, sheet-like structure, the energy density of domain walls decreases more slowly with the cosmic expansion than that of both radiation and matter. Consequently, unless they are sufficiently light, unstable, or otherwise suppressed, domain wall networks would eventually dominate the total energy density of the Universe. Such a scenario is ruled out by observations of the cosmic microwave background \cite{Sousa:2015cqa,Caloni:2026dyu} and by the requirement that it does not spoil the successful predictions of standard cosmology. This constitutes the well-known domain wall problem \cite{Zeldovich:1974uw,Vilenkin:1984ib}, leading to stringent constraints on particle physics models with spontaneously broken discrete symmetries. 

Several mechanisms have been proposed to resolve this issue by ensuring that domain wall networks decay sufficiently early. Two of the most studied are vacuum population bias and explicit symmetry breaking \cite{Coulson:1995nv,Larsson:1996sp,Hindmarsh:1996xv,Avelino:2008qy,Correia:2014kqa,Correia:2018tty, Li:2025gld}. In the former, an initial asymmetry in the occupation probabilities of the vacua leads the network to preferentially evolve toward the dominant vacuum, ultimately resulting in the decay of the domain wall network. In the latter, a small difference in vacuum energy density between nearly degenerate vacua generates a volume pressure that accelerates walls toward the higher-energy-density vacuum, causing its regions to shrink and disappear, thereby driving the decay of the network.

Beyond these standard scenarios, it has been shown that more subtle forms of asymmetry in the underlying field theory can also induce domain wall decay, even when the vacuum energy densities are exactly degenerate. In models with multiple coupled scalar fields, for example, differences in wall tensions can generate biases that destabilize the network \cite{Avelino:2008mh}. Similarly, in single-field models, asymmetries in the shape of the potential near its maximum have been identified as a possible source of biased domain wall network evolution \cite{Krajewski:2021jje}. In addition, scattering and absorption of scalar radiation by domain walls can also influence their evolution \cite{Romanczukiewicz:2017hdu, Forgacs:2008az}. Additional effects, such as plasma-induced friction \cite{Rubin:2001yw, Blasi:2022ayo}, can further modify the wall dynamics and delay or accelerate the annihilation process (see also \cite{Filippov:2025dpb,Kitajima:2023kzu}). 

In this work, we revisit the origin of biased domain wall evolution in theories with degenerate vacua and an asymmetric scalar field potential, with particular focus on scenarios in which the scalar field mass depends on the vacuum state.  We investigate the dynamics of domain walls in such theories in $1+1$ and $2+1$ dimensions, combining analytical arguments with numerical simulations. We study how the vacuum dependence of the scalar field mass affects the scalar radiation emitted by accelerating domain walls, and how the corresponding backreaction modifies their dynamics. Finally, we discuss the implications of this effect for the cosmological evolution of domain wall networks.

Throughout this work, we adopt the metric signature $[-,+,+,+]$ and natural units in which $c=1$. We use the Einstein summation convention, such that repeated indices are summed over their full range. Greek indices run over spacetime components ${0,1,2,3}$, while Latin indices run over spatial components ${1,2,3}$ unless otherwise stated.	

\section{Model Setup}\label{sec:model}

We consider a real scalar field $\phi$ in $(3+1)$-dimensional spacetime with action
\begin{equation}
S = \int d^{4}x \, \sqrt{-g}\,\mathcal{L}\,,
\end{equation}
where $g \equiv \det(g_{\mu\nu})$ and $g_{\mu\nu}$ are the metric components. The scalar field Lagrangian is given by
\begin{equation}
\mathcal{L} = -\frac{1}{2} \nabla_\mu \phi \nabla^\mu \phi - V(\phi)\,,
\label{eq_lagr}
\end{equation}
where $V(\phi) \ge 0$ is a scalar potential with two discrete degenerate minima and $\nabla_\mu$ denotes a covariant derivative with respect to $x^\mu$. The components of the corresponding energy-momentum tensor are
\begin{equation}
T_{\mu\nu} \equiv -\frac{2}{\sqrt{-g}} \frac{\delta(\mathcal L \sqrt{-g})}{\delta g^{\mu \nu}} = \nabla_\mu \phi \nabla_\nu \phi + g_{\mu\nu}\mathcal{L}\,,
\label{eq_EM_tensor}
\end{equation}
and variation of the action with respect to $\phi$ yields the equation of motion
\begin{equation}
\Box \phi \equiv \nabla_\mu \nabla^\mu \phi = \frac{dV}{d\phi}\,.
\label{eq_eom_cov}
\end{equation}

At cosmological scales, the background spacetime is described by a spatially flat Friedmann–Lemaître–Robertson–Walker (FLRW) metric,
\begin{equation}
ds^2 = a^2(\eta)\left(-d\eta^2 + dx^2 + dy^2 + dz^2\right)\,,
\end{equation}
where $(x,y,z)$ are comoving Cartesian coordinates, $a(\eta)$ is the scale factor, and $\eta$ denotes conformal time, defined by $d\eta = dt/a(t)$ with $t$ being the physical time. Equation~\eqref{eq_eom_cov} then becomes
\begin{equation}
\ddot{\phi} + 2 {\mathcal H} \dot{\phi} - \nabla^2 \phi = -a^2\frac{dV}{d\phi}\,,
\end{equation}
where a dot denotes a derivative with respect to the conformal time $\eta$, $\mathcal H \equiv \dot{a}/a = H a$, $H$ is the Hubble parameter, and $\nabla^2$ is the comoving Laplacian. The energy density is then given by
\begin{equation}
\rho = -{T^0}_0 = \frac{1}{2a^2}\left(\dot{\phi}^2 + |\nabla \phi|^2\right) + V(\phi)\,,
\label{eq_energydensity}
\end{equation}
where $\nabla \phi$ denotes the comoving gradient of $\phi$. At the level of microscopic dynamics governing scalar radiation emission by domain walls, the relevant physical length and time scales are typically much smaller than the Hubble scale $H^{-1}$. In this regime, cosmic expansion can be neglected and the spacetime treated as effectively Minkowskian (in this case, one may set $a = 1$, $d\eta = dt$, and $H = 0$ to an excellent approximation).

Although domain wall networks evolve in three spatial dimensions, locally planar configurations depend only on the coordinate normal to the wall. Throughout this work, we shall either employ an effective $(1+1)$-dimensional description, $\phi=\phi(x,t)$, neglecting transverse gradients, or an effective $(2+1)$-dimensional description, $\phi=\phi(x,y,t)$, assuming translational invariance along the $z$-direction. These simplifications capture the essential local dynamics relevant for our analysis and do not affect our main results.

For static configurations in Minkowski spacetime with $\phi=\phi(x)$, the equation of motion reduces to
\begin{equation}
\phi'' = \frac{dV}{d\phi}\,,
\end{equation}
where a prime denotes a derivative with respect to $x$.
Solutions saturating the Bogomolny--Prasad--Sommerfield bound satisfy
\begin{equation}
\phi' =  \pm \sqrt{2V(\phi)}\,.
\label{eq_BPS}
\end{equation}
This equation admits the following integrated implicit general solution:
\begin{equation}
\int_{\phi_0}^{\phi(x)} \frac{d\chi}{\sqrt{2V(\chi)}} = \pm (x - x_0)\,.
\end{equation}
This relation determines the static spatial domain wall profile $\phi=\phi(x)$ for a given potential $V(\phi)$.
The corresponding wall tension (energy per unit transverse area) is
\begin{equation}
\sigma = \int_{-\infty}^{+\infty} \rho(x)\, dx
= \int_{\phi_{\rm L}}^{\phi_{\rm R}} \sqrt{2V(\phi)}\, d\phi\,,
\label{eq_mass_wall}
\end{equation}
where $\phi_{\rm L}$ and $\phi_{\rm R}$ are the asymptotic vacuum values at $x \to \pm\infty$, and the scalar field masses in the left and right vacua are given by
\begin{equation}
m_{\rm L}^2 \equiv \left.\frac{d^2 V}{d\phi^2}\right|_{\phi_{\rm L}} \,, \qquad m_{\rm R}^2 \equiv \left.\frac{d^2 V}{d\phi^2}\right|_{\phi_{\rm R}}\,.
\end{equation}

    \begin{figure}
        \centering
        \includegraphics[width=0.85\textwidth]{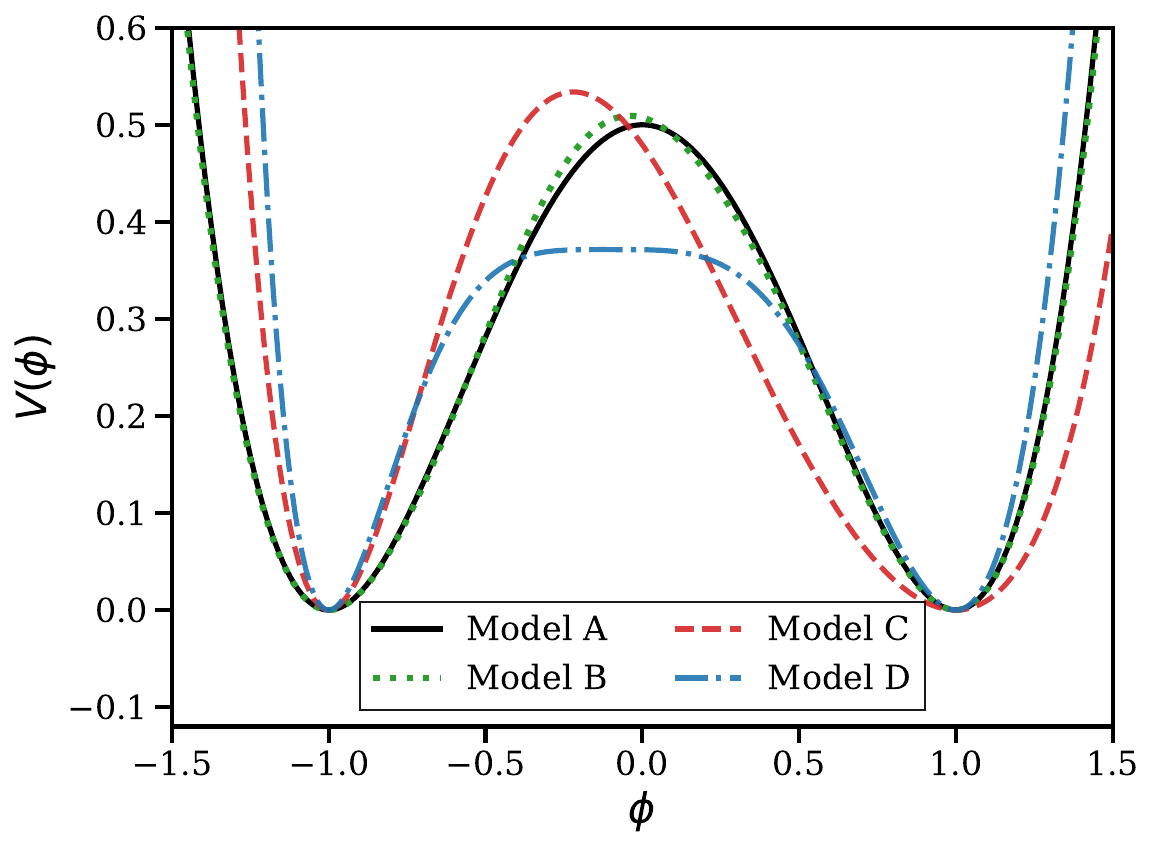}
        \caption{Scalar-field potentials $V(\phi)$ for one symmetric and three asymmetric models. All models share degenerate vacua at $\phi = \pm 1$. The symmetric $\phi^4$ model A (dashed black curve) satisfies $d3V = 0$ and $\Delta m^2 = 0$. Model B (dotted green curve) exhibits a non-zero potential-barrier asymmetry near the local maximum, $d3V \neq 0$, but no mass difference between the vacua, $\Delta m^2 = 0$. Model C (dashed red curve) displays both types of asymmetry, with $d3V \neq 0$ and $\Delta m^2 \neq 0$. Model D (dash-dotted blue curve) is the complementary case of Model B, satisfying $d3V = 0$ and $\Delta m^2 \neq 0$.}
        \label{fig:potentials}
    \end{figure}

To investigate the origin of the dynamical bias induced by asymmetries in the scalar potential, it is useful to disentangle the effects of the vacuum excitation mass splitting, quantified by 
\begin{equation}
\Delta m^2 \equiv |m_{\rm L}^2 - m_{\rm R}^2|\,,
\end{equation}
from the asymmetry of the potential barrier near the local maximum, parameterized by
\begin{equation}
d3V \equiv \left.\frac{d^3 V}{d\phi^3}\right|_{\phi_{\mathrm{max}}}\,.
\end{equation}
Throughout this work, we consider four distinct potential models, constructed to systematically separate these effects (see Fig.~\ref{fig:potentials}).

    \begin{table*}[htbp!]
    \centering
    \caption{Summary of the potential parameters used in Fig.~\ref{fig:potentials} for all four models, together with the associated derived quantities. The normalization parameter $\lambda$ is  tuned to enforce strict domain wall tension degeneracy ($\sigma = 4/3$) across all models.}
    \label{tab:model_parameters}
    \renewcommand{\arraystretch}{1.3}
    \begin{tabular}{lcccccccc}
    \toprule\toprule
    & \multicolumn{4}{c}{\textbf{Potential parameters}} & \multicolumn{4}{c}{\textbf{Derived quantities}} \\
    \cmidrule(lr){2-5} \cmidrule(lr){6-9}
    \textbf{Model} & $b$ & $d$ & $\epsilon$ & $\lambda$ & $m_{\rm L}$ & $m_{\rm R}$ & $\Delta m^2$ & $d3V$ \\
    \midrule
    $\ \ \ \ A$ & --- & --- & --- & $1.00$ & $2.00$ & $2.00$ & $0.00$ & $0.00$ \\
    $\ \ \ \ B$ & --- & --- & $0.03$ & $0.98$ & $1.98$ & $1.98$ & $0.00$ & $3.50$ \\
    $\ \ \ \ C$ & $1.56$ & $1.62$ & --- & $0.45$ & $3.02$ & $1.35$ & $7.29$ & $3.50$ \\
    $\ \ \ \ D$ & $1.23$ & $2.85$ & --- & $0.36$ & $3.56$ & $2.32$ & $7.29$ & $0.00$ \\
    
    \bottomrule\bottomrule
    \end{tabular}
\end{table*}

In all cases, we introduce a dimensionless normalization parameter $\lambda$, which controls the overall amplitude of the potential without affecting the vacuum locations or its shape, and can be used to normalize the domain wall tension. For a consistent comparison across all models, $\lambda$ is chosen so that the domain wall tension is fixed at $\sigma = 4/3$, matching that of the canonical $\phi^4$ model with $\lambda = 1$. The units used throughout this paper are defined by fixing $\sigma = 4/3$, $|\phi_{\rm L}| = |\phi_{\rm R}| = 1$, and $c = 1$ (as introduced earlier). This choice completely specifies the system of units adopted in this work. The parameters used in Fig.~\ref{fig:potentials} for all four models, together with the associated derived quantities, are summarized in Table~\ref{tab:model_parameters}.

\subsection{Model A ($d3V=0 \,, \ \Delta m^2 =0$)}

As a control case, we consider the standard $\phi^4$ model defined by the potential
\begin{equation}
V_{\rm A}(\phi) = \frac{\lambda}{2}(\phi^2 - 1)^2 \,.
\end{equation}
In this model, the potential is an even function of $\phi$, implying that $d3V=0$ and that the scalar field masses are identical in the two degenerate vacua ($m_{\rm L} = m_{\rm R} = 2$). For $\lambda=1$, integrating Eq.~\eqref{eq_mass_wall} yields the domain wall tension $\sigma = 4/3$. The corresponding potential is represented by the solid black curve in Fig.~\ref{fig:potentials}.

\subsection{Model B ($d3V \neq  0 \,, \ \Delta m^2  = 0$)} 

The potential of model B is constructed by introducing a localized deformation on  $V_{\rm A}$:
\begin{equation}
V_{\rm B}(\phi) = \lambda \left\{ \frac{1}{2}(\phi^2 - 1)^2 + \epsilon \cos^2\!\big(\pi(\phi + 0.2)\big)\, \Big[\Theta(\phi + 0.7) - \Theta(\phi - 0.3)\Big] \right\},
\end{equation}
where $\Theta$ is the Heaviside step function and $\epsilon$ controls the amplitude of the localized deformation. This modification leaves the vacuum structure unchanged, thereby preserving the mass degeneracy of model A, $\Delta m^2 = 0$, while introducing an asymmetry in the potential barrier near the local maximum, such that $d3V \neq 0$. The resulting potential corresponds to the dotted green curve in Fig.~\ref{fig:potentials}.

\subsection{Model C ($d3V \neq 0 \,, \ \Delta m^2  \neq 0$)}

In order to define model C we start by adopting a polynomial potential form considered in Ref.~\cite{Krajewski:2021jje}, defined via an auxiliary field $\varphi$:
\begin{equation}
\frac{d U_{\rm C}}{d \varphi} (\varphi) 
= \lambda (\varphi + 1)(\varphi - b)\varphi \left[(\varphi - d)^2 + 1 \right].
\end{equation}
For simplicity, we define the potential $U_{\rm C}(\varphi)$ through its analytical derivative, as the explicit form of the potential is rather lengthy. The vacuum states of this auxiliary potential are located at $\varphi_{\rm L} = -1$ and $\varphi_{\rm R} = b$. The parameters $b$ and $d$ are tuned so that the vacuum energies remain exactly degenerate [$U_{\rm C}(\varphi_{\rm L}) = U_{\rm C}(\varphi_{\rm R})$]. This construction generally yields $\Delta m^2 \neq 0$ and $d3 V \neq 0$.

In order to keep the vacua at $\phi= \pm 1$, we perform the following affine transformation
\begin{equation}
\phi = \frac{2\varphi - b+1}{b+1}\,, \label{affine1}
\end{equation}
which maps $\varphi \in [-1,b]$ to $\phi \in [-1,1]$. To ensure that $m_{\rm L}$ and $m_{\rm R}$ are invariant under this rescaling, the potential must be redefined as
\begin{equation}
V_{\rm C}(\phi) = \left(\frac{2}{b+1}\right)^2\,U_{\rm C}(\varphi)\,. \label{affine2}
\end{equation}
This model is illustrated by the dashed red curve in Fig.~\ref{fig:potentials}.

\subsection{Model D ($d3V = 0 \,, \ \Delta m^2  \neq 0$)}

Model D considers a modified potential based on model C, but with an additional $\varphi^3$ factor that suppresses the asymmetry of the potential barrier near the local maximum:
\begin{equation}
\frac{d U_{\rm D}}{d \varphi}(\varphi) 
=  \lambda (\varphi + 1)(\varphi - b) \varphi^3 \left[(\varphi - d)^2 + 1 \right]. \label{dVC}
\end{equation}
By construction, model D satisfies $d3V = 0$, while allowing for a non-zero mass difference between the vacua ($\Delta m^2 \neq 0$). The potential $V_{\rm D}(\phi)$ is obtained by integrating Eq.~\eqref{dVC}, then applying the affine transformation of Eq.~\eqref{affine1} and the rescaling by $\left[2/(b+1)\right]^2$ in Eq.~\eqref{affine2}. The resulting $V_{\rm D}(\phi)$ corresponds to the dash-dotted blue curve in Fig.~\ref{fig:potentials}.

\section{Origin of the Domain Wall Rocket Effect}\label{sec:rocket}

Consider a planar domain wall in Minkowski spacetime oriented perpendicular to the $x$-direction, so that the dynamics is effectively reduced to $1+1$ dimensions. Note that although a $(1+1)$-dimensional kink has no area, we consider here a planar wall in 3+1 dimensions, with all extensive quantities defined per unit transverse area. In the presence of a moving domain wall, it is convenient to introduce a collective coordinate $X(t)$ that describes its position. The wall is then centered at $x = X(t)$, and its motion is encoded in the time dependence of this coordinate. We define
\begin{equation}
\xi = x - X(t)\,,
\end{equation}
which measures the distance from the instantaneous position of the wall.

Far from the wall, the field approaches a vacuum configuration and the dynamics become linear. In this regime, the field can be decomposed into plane-wave modes,
\begin{equation}
\phi_k(x,t) = A_k \cos(\omega_k t - kx - \delta_k)\,,
\end{equation}
where the sign of $k$ encodes the direction of propagation. These modes satisfy the dispersion relation
\begin{equation}
\omega_k^2 = k^2 + m^2(\xi)\,,
\label{eq_dispersion}
\end{equation}
where $\omega_k$ denotes the angular frequency associated with the wavenumber $k$ and the scalar field mass is defined piecewise as
\begin{equation}
m(\xi) =
\begin{cases}
m_{\rm L}\,, & \xi < 0\,, \\
m_{\rm R}\,, & \xi > 0\,.
\end{cases}
\end{equation}
This sharp transition is an idealization of a more gradual spatial variation of the mass across the domain wall. Only modes satisfying $\omega_k \ge m(\xi)$ correspond to propagating radiation (otherwise, the excitations are evanescent and decay exponentially).

A non-zero $\Delta m^2$ has direct consequences for domain wall dynamics through the kinematic constraint given in \eqref{eq_dispersion}. For oscillation frequencies in the window $m_{\rm R} < \omega < m_{\rm L}$, radiation emitted by the oscillating wall can propagate freely into the right vacuum, while remaining evanescent in the left one. This asymmetric filtering of modes results in an imbalance in the outgoing radiation flux across the wall, giving rise to a net recoil force per unit area on the domain wall. Although the spectral distribution of the emitted modes is largely determined by the dynamics of the wall, their propagation is controlled by the vacuum structure, which acts as a kinematic filter selectively allowing or suppressing propagation depending on whether $\omega_k \gtrless m(\xi)$.

To quantify this effect, we consider the energy-momentum tensor of the scalar field,
\begin{align}
T^{00} &= \frac{1}{2}\dot{\phi}^2 + \frac{1}{2}\phi'^2 + V(\phi)\,, \\
T^{01} &= -\dot{\phi}\,\phi'\,, \\
T^{11} &= \frac{1}{2}\dot{\phi}^2 + \frac{1}{2}\phi'^2 - V(\phi)\,.
\end{align}
Averaging over many oscillation periods removes rapidly oscillating terms, yielding
\begin{align}
\langle T^{00}_k \rangle &= \frac{1}{2} A_k^2 \omega_k^2\,, \\
\langle T^{01}_k \rangle &= \frac{1}{2} A_k^2 \omega_k k\,, \\
\langle T^{11}_k \rangle &= \frac{1}{2} A_k^2 k^2\,.
\end{align}

Using the dispersion relation [Eq.~\eqref{eq_dispersion}], the  energy and linear momentum fluxes carried by a single mode can be written as
\begin{align}
 \mathcal{F}_{E,k}  & \equiv \langle T^{01}_k \rangle = \frac{1}{2} A_k^2 \omega_k \sqrt{\omega_k^2 - m^2(\xi)}\,\Theta\left[\omega_k - m(\xi)\right]\,, \\
\mathcal{F}_{p,k}  &\equiv \langle T^{11}_k \rangle = \frac{1}{2} A_k^2 \left[\omega_k^2 - m^2(\xi)\right]\,\Theta\left[\omega_k - m(\xi)\right]\,,
\end{align}
where $\Theta$ is the Heaviside step function. On macroscopic timescales, interference terms between distinct Fourier modes average to zero, so that the total energy and momentum fluxes can be obtained by integrating the contributions of the individual modes over the spectral distribution,
\begin{equation}
 \mathcal{F}_E  = \int \frac{dk}{2\pi}\,\mathcal{F}_{E,k} \,, \qquad 
 \mathcal{F}_p  = \int \frac{dk}{2\pi}\,  \mathcal{F}_{p,k} \,.
\end{equation}

A larger $m$ increases the propagation threshold and reduces the wavenumber $k$ of the transmitted modes. Integrating over all modes yields a net flux asymmetry, favoring emission into the vacuum with the lower scalar field mass. The resulting net force per unit area acting on the wall is
\begin{equation}
F_{\mathrm{net}}(t) = \mathcal{F}_{p, \rm L}(t) - \mathcal{F}_{p, \rm R}(t)\,.
\end{equation}

Defining the accumulated momentum per unit area
\begin{equation}
P(t) \equiv \int_0^t F_{\mathrm{net}}(t')\, dt',
\end{equation}
the velocity can be written as
\begin{equation}
v(t) = \frac{P(t)}{\sqrt{\sigma^2 + P(t)^2}}. \label{eq_velocity_th}
\end{equation}
This shows how asymmetries in the dispersion relation may contribute to an imbalance in the energy and momentum fluxes, inducing a recoil that drives the wall toward the vacuum with the larger scalar field mass, thereby dynamically favoring the expansion of the vacuum with the smaller scalar field mass.

\section{Rocket Effect in $1+1$ Dimensions}

To develop a clear physical understanding of the rocket effect, we begin by analyzing the $(1+1)$-dimensional dynamics of a perturbed planar domain wall in Minkowski spacetime.

\subsection{Simulation setup and initialization}

We solve the full non-linear equation of motion [Eq.~\eqref{eq_eom_cov}] on a discretized one-dimensional lattice in order to track domain wall relaxation, scalar radiation emission, and the resulting domain wall recoil. The spatial domain has physical length $L=400$ and is discretized with $N=10^6$ grid points, corresponding to a resolution $\Delta x = 0.004$. Time evolution is performed with a step size $\Delta t = 4\times10^{-4}$. Spatial lattice sites are labeled by $i$ and discrete time steps by $n \ge 0$, such that $t^{[n]} = n\Delta t$. The numerical solver is optimized using Just-In-Time (JIT) compilation. For a comprehensive overview of the finite-difference methods, integration schemes, absorbing boundary conditions and numerical data analysis, see~\cite{newman2013computational, press2007numerical, thomas1995numerical}.

Time integration is carried out using a staggered leapfrog scheme. To incorporate both initial relaxation and absorbing boundary conditions without compromising the symplectic structure of the bulk evolution, we introduce a position-dependent damping coefficient $\Gamma_i$:
\begin{align} 
    \dot{\phi}_{i}^{[n+1/2]} &= \frac{\dot{\phi}_{i}^{[n-1/2]} + \Delta t \left[ (\nabla^2 \phi)_{i}^{[n]} - \left.\frac{\partial V}{\partial \phi}\right|_{\phi_{i}^{[n]}} \right]}{1 + \frac{1}{2} \Gamma_{i} \Delta t}\label{dotphi1d}\,,  \\ 
    \phi_{i}^{[n+1]} &= \phi_{i}^{[n]} + \Delta t \, \dot{\phi}_{i}^{[n+1/2]}\,.
\end{align}
In the bulk simulation region during data acquisition, $\Gamma_i=0$, ensuring Hamiltonian evolution. Nonzero damping is introduced only in controlled regions, as described below.

First, to eliminate high-frequency artifacts introduced by interpolating continuous initial profiles onto the lattice, we perform a pre-relaxation phase. The system is evolved for $1500$ time steps with a uniform damping $\Gamma_i=5.0$, which efficiently removes spurious numerical excitations. The damping is then switched off, and the simulation is reset to $t=0$. Second, to suppress spurious reflections from the boundaries, we implement absorbing boundary conditions using a sponge layer. The lattice is extended by $15\%$ beyond the physical domain, and within this padding region we set $\Gamma_i=\gamma(x_i)$, where $\gamma(x)$ increases smoothly toward the boundary following a cubic profile. This procedure effectively damps outgoing radiation.

Spatial derivatives are computed using a second-order central difference stencil:
\begin{equation}
    (\nabla^2 \phi)_i^{[n]} = \frac{\phi_{i+1}^{[n]} - 2\phi_i^{[n]} + \phi_{i-1}^{[n]}}{\Delta x^2}\,.
\end{equation}
Adequate resolution of the domain wall core is essential to avoid lattice artifacts such as pinning. The domain wall thickness scales as $w \sim m^{-1}$, where $m$ is the vacuum excitation mass. Using the vacuum with larger scalar field mass in our models ($m = 3.56$), we obtain a minimum thickness $w \simeq 0.28$, corresponding to $\simeq 70$ grid points at our resolution. Independent measures based on the full width at half maximum of the energy density and the geometric width $\delta_{\text{geo}} = (\phi_{\rm R} - \phi_{\rm L}) / |\nabla\phi|_{\max}$ yield even higher effective resolutions, confirming that the dynamics are well resolved. 

The unperturbed wall profile $\phi_{\text{static}}(x)$ is determined analytically for model A, while for the asymmetric models (B, C and D) it is generated by numerically integrating Eq.~\eqref{eq_BPS}. The resulting solution is used to construct a high-precision cubic spline interpolant, which is then evaluated across the lattice. To excite the domain wall, we perturb the static configuration using the shape mode profile of the $\phi^4$ kink:
\begin{equation}
    \eta_{\rm A}(x) = \sqrt{\frac32}\sech^2(x-x_0)\,\sinh(x-x_0)\,,
\end{equation}
The initial condition is
\begin{equation}
    \phi(x,0) = \phi_{\text{static}}(x) + C \eta_{\rm A}(x)\,.
\end{equation}
where $C$ is a normalization constant. $C$ is determined numerically for each model using a numerical root-finding procedure designed to attain a fixed target value for the domain wall energy per unit area. This procedure provides a consistent and standardized prescription for the initial energy injection across all models.

\begin{figure}[htbp]
    \centering
    \includegraphics[width=0.85\textwidth]{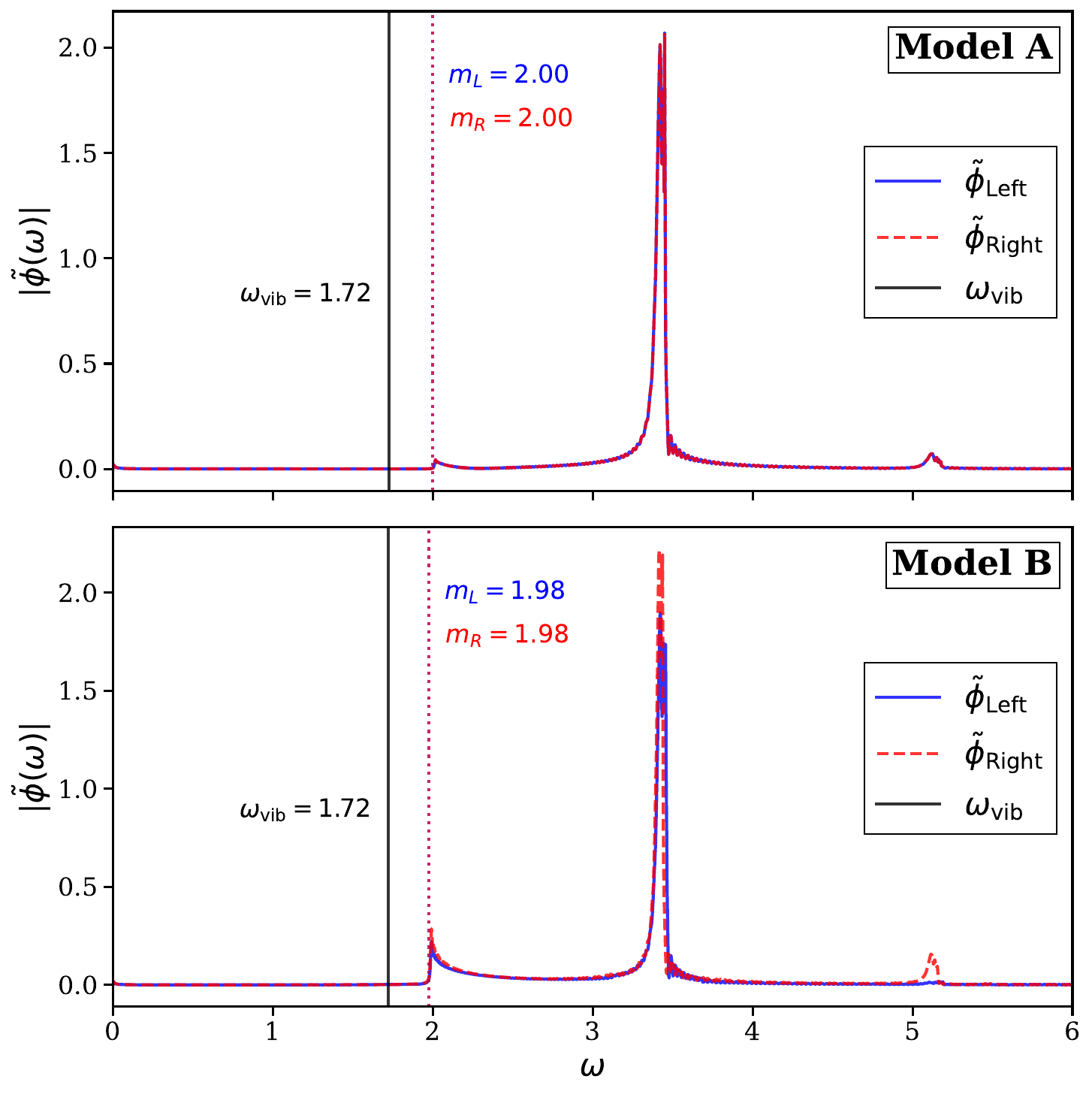}
    \caption{Amplitude spectra of the emitted scalar radiation, $|\tilde{\phi}(\omega)|$, measured by far-field probes in the left (solid blue) and right (dashed red) vacua. Vertical dotted lines indicate the mass thresholds $m_{\rm L}$ and $m_{\rm R}$ for the propagation of scalar radiation, while the solid black line marks the fundamental shape mode frequency $\omega_{\rm vib}$. Model A exhibits symmetric emission, whereas model B shows only a weak asymmetry, reflected in a slightly reduced peak amplitude in the left vacuum.}
    \label{fig:fft_spectra_AB}
\end{figure}

\begin{figure}[htbp]
    \centering
    \includegraphics[width=0.85\textwidth]{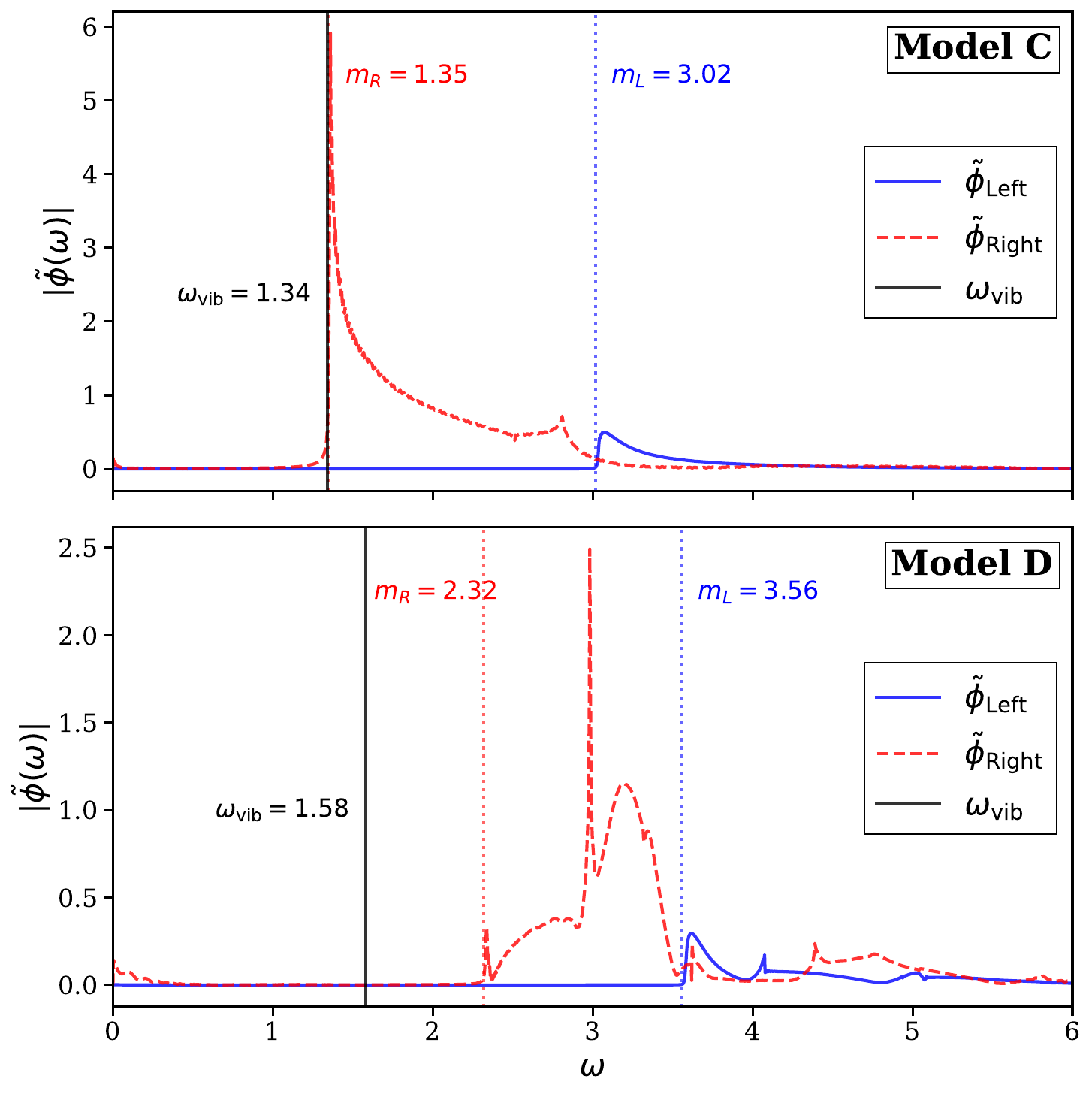}
    \caption{Same as Fig.~\ref{fig:fft_spectra_AB}, but for the mass-asymmetric models (C and D). In these models, most of the power lies in the range $m_{\rm R} < \omega < m_{\rm L}$, which leads to a strong suppression of the propagation of scalar radiation into the left vacuum and results in predominantly rightward emission.}
    \label{fig:fft_spectra_CD}
\end{figure}

To characterize the internal dynamics of the walls and their coupling to radiation, we first extract the fundamental shape mode vibrational frequency $\omega_{\rm vib}$. Each model possesses its own intrinsic shape mode $\eta_{\text{true}}(x)$, but we define
\begin{equation}
    \phi_{\rm proj}(t) = \frac{\int \phi(x,t)\,\eta_{\rm A}(x)\,dx}{\int \eta_{\rm A}^2(x)\,dx}\,,
\end{equation}
by projecting the field $\phi(x,t)$ onto the $\phi^4$ shape mode profile $\eta_{\rm A}$. Although $\eta_{\rm A}(x)$ is an exact eigenmode only for model A, it is employed as a fixed spatial filter across all models. The resulting projection provides a measure of the vibrational dynamics of the wall and exhibits a distinct spectral peak associated with the model-dependent shape mode, from which $\omega_{\rm vib}$ is extracted via Fourier analysis of $\phi_{\rm proj}(t)$.

We independently confirmed the value of $\omega_{\rm vib}$ for each model through a linear stability analysis of the corresponding static solutions. The resulting eigenvalue problem is solved using the Lanczos algorithm, an iterative Krylov-subspace method well suited to large sparse Hermitian operators, which efficiently extracts the lowest-lying eigenvalues and their associated eigenmodes (see \cite{saad2011numerical} for more details).

The scalar radiation spectrum is obtained by monitoring the time evolution of the scalar field in the left and right far-field regions (at $x=5$ and $x=395$). The Fourier transform $\tilde{\phi}(\omega)$ of the resulting discrete time series is computed using a Fast Fourier Transform (FFT).

\subsection{Radiation Spectra and Emission Bias}

\begin{figure}[htbp]
    \centering
    \includegraphics[width=1.0\textwidth]{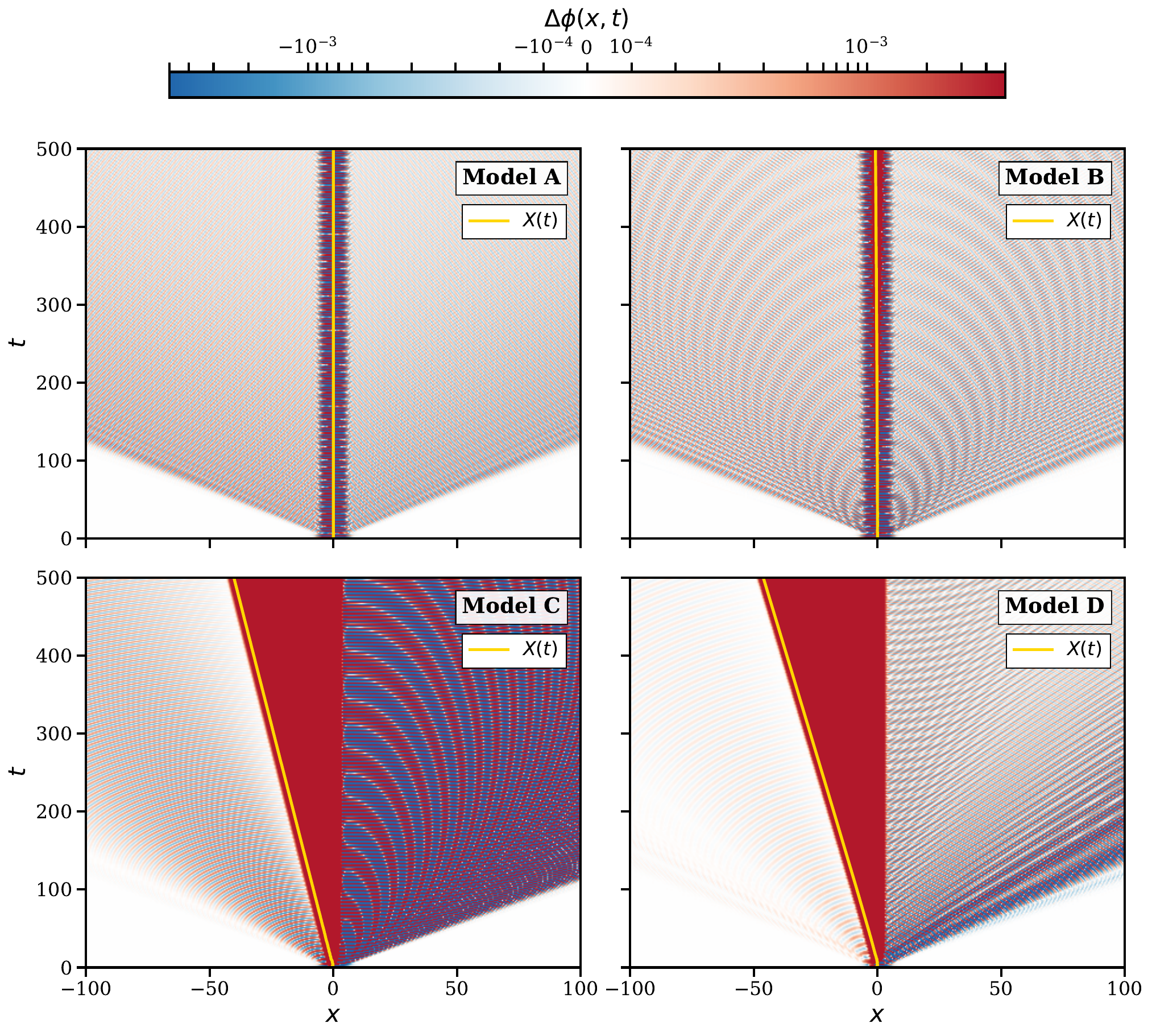} 
    \caption{Spacetime evolution of the scalar radiation field $\Delta\phi(x,t)$ for models A–D, centered on the initial defect position. The view is restricted to $x \in [-100,100]$ and $t \in [0,500]$ to highlight the wall trajectory and primary emission fronts. A symmetrically saturated logarithmic color scale (at $\pm 5\times10^{-3}$) enhances faint radiation, with saturated regions near the core indicating larger amplitudes. Model A exhibits symmetric emission and a stationary wall (gold line), while model B shows only a negligible asymmetry at this scale. In contrast, models C and D display strongly asymmetric emission, preferentially directed into the right vacuum, which induces a net recoil of the wall toward the opposite side.}
    \label{fig:heatmap_2x2}
\end{figure}

Figure~\ref{fig:fft_spectra_AB} shows the amplitude spectra of the emitted scalar radiation, $|\tilde{\phi}(\omega)|$, measured by far-field probes over the simulation time interval $t \in [0, 2000]$, in the left (solid blue) and right (dashed red) vacua. The spectra are obtained from the Fourier transform of the scalar-field time series recorded at the corresponding probe locations. Vertical dotted lines indicate the mass thresholds $m_{\rm L}$ and $m_{\rm R}$, which delimit the onset of the continuum of propagating radiation in the respective vacua. The solid black line denotes the fundamental shape mode frequency $\omega_{\rm vib}$.

In the mass-degenerate models (A and B, with $\Delta m^2$ = 0), the shape mode frequency lies below the continuum threshold of the linear spectrum. Energy loss occurs only through nonlinear interactions, in particular via higher harmonics that enter the continuum, leading to symmetric scalar radiation emission in model A or approximately symmetric emission in model B. This behaviour is clearly reflected in the far-field spectra, shown in Fig.~\ref{fig:fft_spectra_AB}. In model A, the radiation is symmetric and dominated by the second harmonic, with only subleading contributions from higher modes. Model B displays a weak asymmetry: although the mass thresholds are identical, the asymmetry of the vacuum barrier near the local maximum is responsible for a slight imbalance in the emitted power between the two sides, reflected in a slightly reduced peak amplitude in the left vacuum compared to the right one.

Figure~\ref{fig:fft_spectra_CD} shows the same analysis as in Fig.~\ref{fig:fft_spectra_AB}, but for the mass-asymmetric models (C and D). In these models, most of the spectral power is concentrated in the frequency window $m_{\rm R} < \omega < m_{\rm L}$, which lies above the mass threshold of the right vacuum but below that of the left vacuum. As a consequence, scalar radiation can propagate efficiently into the right vacuum, while it becomes evanescent in the left vacuum. This leads to a strong suppression of radiation transport into the left vacuum and results in predominantly rightward emission.

This is confirmed in Fig.~\ref{fig:fft_spectra_CD}, where models with $\Delta m^2 \neq 0$ (C and D) exhibit strongly asymmetric energy loss. In both cases, since $m_{\rm R} < m_{\rm L}$, the spectra are strongly biased toward the right vacuum, whereas the left vacuum exhibits only a weak near-threshold accumulation with negligible support from propagating modes. In model C, the shape mode lies essentially at the propagation threshold set by the smaller vacuum mass $m_{\rm R}$, making it marginally bound at best. As a result, it fails to remain a fully localized excitation and instead efficiently couples to the continuum, radiating predominantly into the right vacuum. In model D, by contrast, the fundamental mode remains localized within the wall. However, its first harmonic (identified as the dominant peak of the red-dashed spectrum) enters the propagating band of the right vacuum while remaining kinematically forbidden in the left one. This mismatch between the allowed spectral bands effectively acts as a directional filter for the radiation.

Overall, Fig.~\ref{fig:fft_spectra_CD} shows that vacuum mass thresholds act as sharp filters for scalar radiation: only modes above the continuum threshold can propagate, forcing radiation into the vacuum with smaller scalar field mass and thereby naturally generating anisotropic emission.

Figure~\ref{fig:heatmap_2x2} displays the spacetime evolution of the scalar radiation field, defined as 
\begin{equation}
\Delta\phi(x,t) \equiv \phi(x,t) - \phi_{\text{static}}[x - X(t)]    \,,
\end{equation}
together with the corresponding domain wall trajectories for all four models. The plots are restricted to the region $x \in [-100,100]$ and $t \in [0,500]$ in order to better resolve the wall motion and the primary radiation fronts. The symmetrically saturated logarithmic color scale (at $\pm 5\times10^{-3}$) enhances low-amplitude radiation, while saturated regions near the wall core correspond to larger field amplitudes. As shown in Fig.~\ref{fig:heatmap_2x2}, models A and B, both with $\Delta m^2 = 0$, exhibit radiation that is emitted symmetrically or nearly symmetrically into the two vacua. In particular, the spacetime patterns in model B remain qualitatively similar to those of model A, indicating that the asymmetry of the potential barrier alone produces only a weak effect on the radiation profile at the scales displayed in the figure. By contrast, the lower panels of Fig.~\ref{fig:heatmap_2x2}, corresponding to models C and D, both with $\Delta m^2 \neq 0$, reveal a pronounced asymmetry in the emitted radiation. In both cases, the radiation is preferentially emitted into the right vacuum, producing an asymmetric momentum flux that drives a recoil of the domain wall toward the opposite side.

\subsection{Wall motion and recoil velocity}

The velocity of the domain wall can be estimated using two complementary methods. In the first method, the spatial position of the domain wall center $X(t)$ is tracked at each measurement step by identifying the zero-crossing of the scalar field $\phi(x,t)$. The corresponding velocity $v(t)$ is then obtained from a smoothed finite-difference derivative of $X(t)$. In the second method, the velocity is obtained from global momentum conservation, as derived in Sec.~\ref{sec:rocket}. Since the total momentum initially vanishes, the wall momentum per unit area satisfies $P_{\rm wall}(t) = -P_{\rm rad}(t)$, where $P_{\rm rad}(t)$ is computed from the bulk momentum density and the net flux through the simulation boundaries,
\begin{equation}
P_{\rm rad}(t) =
\int_{\rm vacua} T^{01}(x,t)\,dx
+ \int_{0}^{t} \left[ T^{11}(x_{\rm right},t') - T^{11}(x_{\rm left},t') \right] dt' \,. 
\end{equation}
where the region designated by `vacua' consists on the spatial region outside a dynamical core cutoff, $|x - X(t)| > R$, where $R = \max(4.0,4w)$ and $w$ denotes the instantaneous root mean square width of the energy density profile. The wall energy per unit area $E_{\rm wall}(t)$ is obtained by integrating $T^{00}$ over the complementary inner core region, $|x - X(t)| \le R$:
\begin{equation}
E_{\rm wall}(t) =
\int_{X(t)-R}^{X(t)+R} T^{00}(x,t)\,dx \,. \label{wall_energy}
\end{equation}
The recoil velocity then follows from momentum–energy consistency,
\begin{equation}
v(t) = -\frac{P_{\rm rad}(t)}{E_{\rm wall}(t)} \,. \label{velocity_theoretical}
\end{equation}
The two determinations are found to be in excellent agreement, with asymptotic velocities agreeing within an error below $0.4\%$. 

\begin{figure}[htbp]
    \centering
    \includegraphics[width=0.85\textwidth]{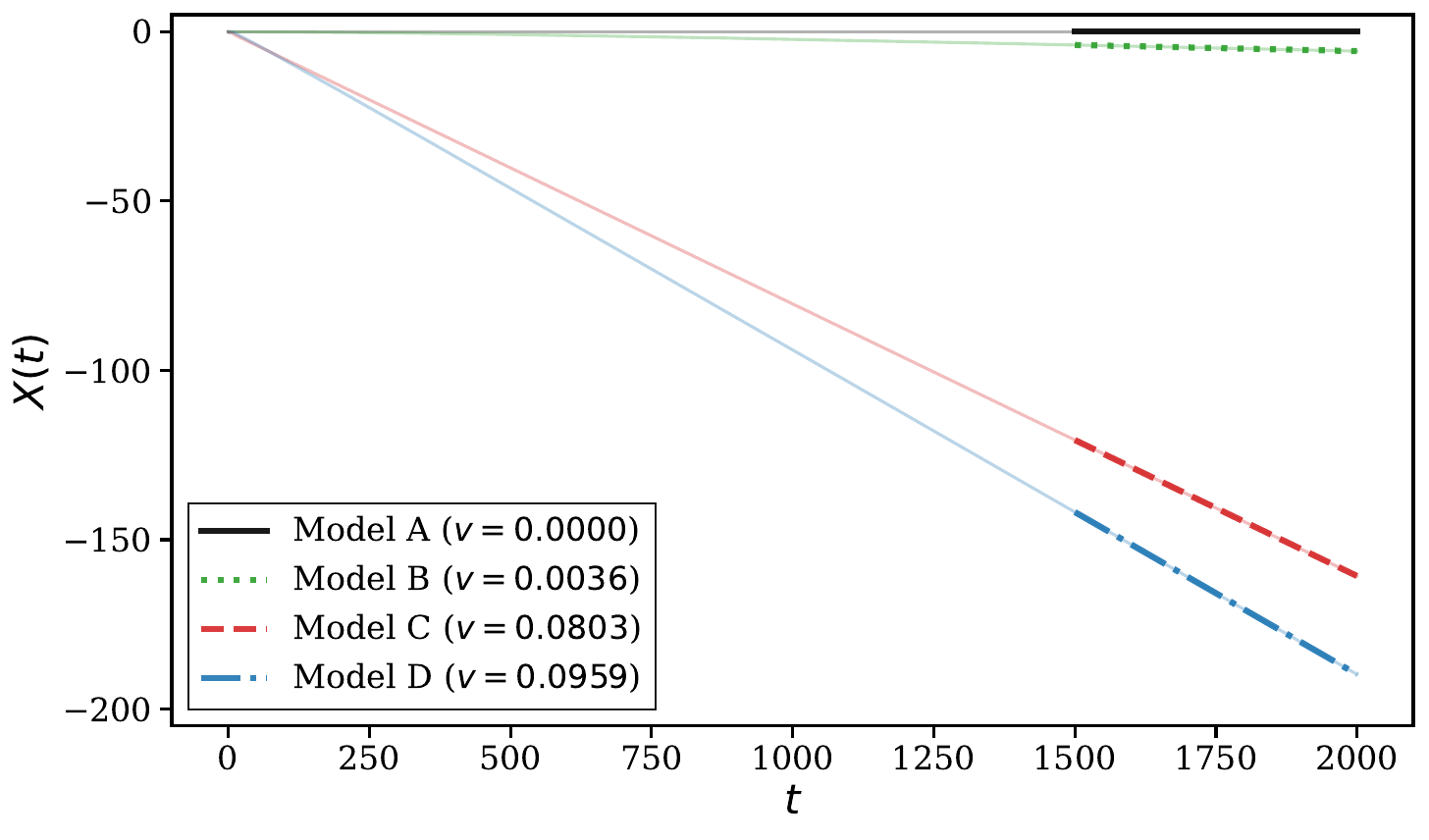}
    \caption{Spatial trajectories $X(t)$ of the domain walls. The extracted terminal velocities (shown in the legend) indicate that domain walls in the models with a vacuum mass differential (models C and D) develop a significant velocity drift and displacement growth. In contrast, domain walls in the skewed but mass-symmetric model B exhibit a much smaller velocity drift and displacement growth, while those in the baseline model A remain stationary throughout the evolution.}
    \label{fig:terminal_vel}
\end{figure}

Figure~\ref{fig:terminal_vel} shows the spatial trajectories $X(t)$ of the domain walls, together with the corresponding extracted terminal velocity fits. As expected, the velocity vanishes for Model A, consistently with its symmetric radiation emission. Models C and D exhibit the largest spatial drifts and terminal velocities, with $v = -0.080$ and $v = -0.096$, respectively. Notably, Model D displays the largest absolute velocity among all cases considered, despite having $d3V = 0$. In contrast, only a residual drift is observed in Model B, with $v = -0.004$, as expected given the vanishing mass difference between the vacua ($\Delta m^2 = 0$ in Model B). The close agreement between Models C and D, in sharp contrast with Models A and B, indicates that the dominant contribution to the rocket effect arises from the vacuum dependence of the scalar field mass, while the impact of a non-zero $d3V$ appears subdominant in comparison.

To quantify the efficiency of the recoil mechanism, we compare the kinetic energy of the wall at terminal velocity with the initially injected energy. For the mass-asymmetric models, the kinetic energy of the accelerated wall accounts for only a small fraction of the injected energy: approximately $2\%$ for model C and $3\%$ for model D. This reflects the nature of the rocket effect: the wall's acceleration is driven by the asymmetry of the emitted radiation, and hence by its net momentum flux, rather than by the amount of energy radiated.

\section{Rocket Effect in $2+1$ Dimensions}

Here we consider the $(2+1)$-dimensional dynamics of a perturbed planar domain wall in Minkowski spacetime. The additional embedding direction endows the wall with geometric degrees of freedom absent in $(1+1)$ dimensions, allowing for curvature and time-dependent undulations of its worldvolume. This richer dynamical setting enables us to investigate the interplay between anisotropic radiation pressure and the evolving wall geometry, providing a stringent test of the robustness of the recoil mechanism.

\subsection{Simulation Setup and Initialization}

The physical domain is discretized on a two-dimensional lattice of size $L \times L$ with $L=30$, using an $N \times N$ grid with $N=2048$ points. This corresponds to a spatial resolution of $\Delta x = 0.0146$, which is sufficient to resolve the wall thickness across all models; the thinnest wall ($w \simeq 0.28$) is thus resolved by approximately $19$ grid points. The time step is chosen to satisfy the Courant–Friedrichs–Lewy (CFL) stability condition, with $\Delta t = 0.2\,\Delta x$.

To prevent grid anisotropy from artificially distorting the propagation of transverse waves, the spatial Laplacian is computed using a 9-point isotropic stencil:
\begin{equation}
(\nabla^2 \phi)_{i,j}^{[n]} = \frac{4 \Sigma_{\text{ortho}} + \Sigma_{\text{diag}} - 20\phi_{i,j}^{[n]}}{6\Delta x^2},
\end{equation}
where $\Sigma_{\text{ortho}}$ denotes the sum over the four nearest (orthogonal) neighbours and $\Sigma_{\text{diag}}$ denotes the sum over the four next-nearest (diagonal) neighbours.

Time evolution is carried out using the staggered leapfrog integration scheme introduced for the $(1+1)$-dimensional simulations [see Eq.~\eqref{dotphi1d}]. To prevent artificial reflections over the course of the simulation, the $x$-axis boundaries employ the $15\%$ cubic sponge layer and radiation conditions as in the $(1+1)$-dimensional setup, absorbing outgoing scalar waves, while periodic boundary conditions are imposed along the $y$-axis.

To probe scalar radiation emission from dynamical domain walls, we initialize the system with a transverse spatial wiggle. This is achieved by applying a periodic spatial displacement to the static profile,
\begin{equation}
    \phi(x,y,t=0) = \phi_{\text{static}}\left[x - x_0 - A_{\text{wiggle}} \sin(k_y y)\right],
\end{equation}
where $x_0$ denotes the initial wall position (before the perturbation) along the $x$-axis, $A_{\text{wiggle}} = 0.4$ is the amplitude of the transverse spatial wiggle, $k_y = 2\pi m / L$ and the static profile is obtained analogously to the $(1+1)$-dimensional case. Since the wall attains maximal extension twice per oscillation cycle, the total wall length exhibits a modulation at frequency $2\omega_{\text{wiggle}}$. This modulation acts as an effective driving force for the internal vibrational dynamics of the domain wall, inducing periodic stretching and compression at the frequency $\omega_{\rm vib}=2\omega_{\rm wiggle}$ (see \cite{Blanco-Pillado:2022rad} for more details). To target an internal vibration with $\omega_{\text{vib}} = 2.52$, which lies between the mass thresholds of the mass-asymmetric models, we choose $m=6$ full wavelengths within the simulation domain.

Before data collection begins, the initialized grid undergoes a strongly damped pre-relaxation phase ($\Gamma_i = 5.0$, 1500 time steps), as in the $(1+1)$-dimensional case. This stage is performed on decoupled $(1+1)$-dimensional horizontal strips by temporarily neglecting transverse derivatives ($\partial\phi/\partial y$). This suppresses interpolation-induced high-frequency noise while preventing relaxation of the imposed curvature. Once each strip has relaxed toward the kink profile, damping is switched off, transverse couplings are restored, and the evolution is reset to $t=0$ for the full $(2+1)$-dimensional dynamics.

\subsection{Radiation Spectra and Emission Bias}

Scalar radiation is monitored using far-field probes located at $x=3$ and $x=27$ over the simulation time interval $t\in \left[ 0, 240 \right]$. For each probe, the field is averaged over the transverse direction,
\begin{equation}
\bar{\phi}(x,t)=\frac{1}{L}\int_0^L \phi(x,y,t)\,dy,
\end{equation}
and Fourier transformed in time to obtain the radiation spectrum $|\tilde{\phi}(\omega)|$. This effectively filters transverse excitations and isolates the longitudinal radiation component.

\begin{figure}[htbp]
    \centering
    \includegraphics[width=0.85\textwidth]{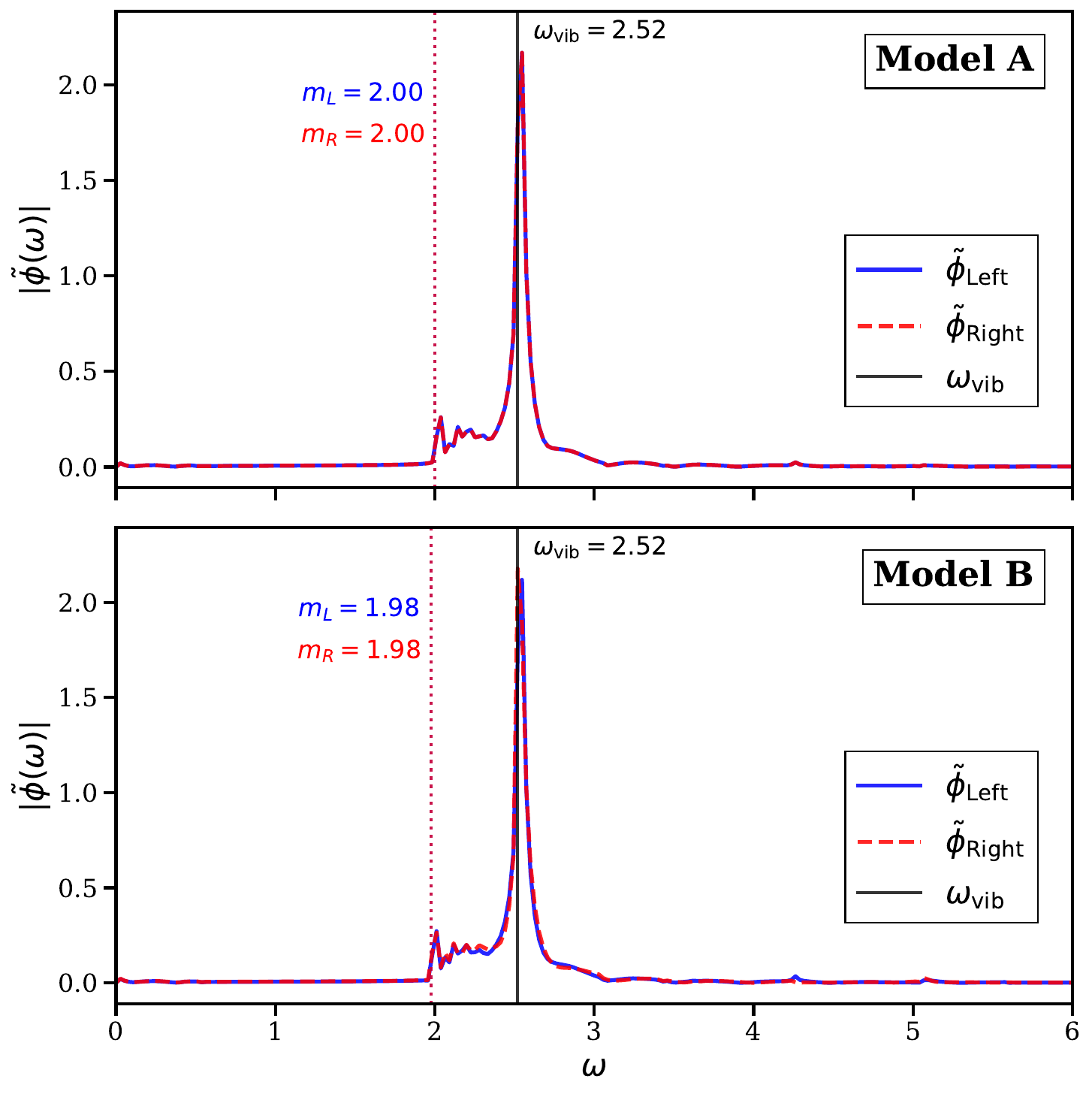}
    \caption{Amplitude spectra of the strip-averaged emitted scalar radiation, $|\tilde{\phi}(\omega)|$, measured by far-field probes in the left (solid blue) and right (dashed red) vacua for the mass-symmetric models (A) and (B). Vertical dotted lines indicate the mass thresholds $m_{\rm L}$ and $m_{\rm R}$ for radiation propagation, while the solid black line marks the wall's vibrational frequency $\omega_{\rm vib}$. Model A exhibits symmetric emission, whereas model B shows only a weak asymmetry, reflected in a slightly reduced peak amplitude in the left vacuum.}
    \label{fig:fft_spectra_2D_AB}
\end{figure}

\begin{figure}[htbp]
    \centering
    \includegraphics[width=0.85\textwidth]{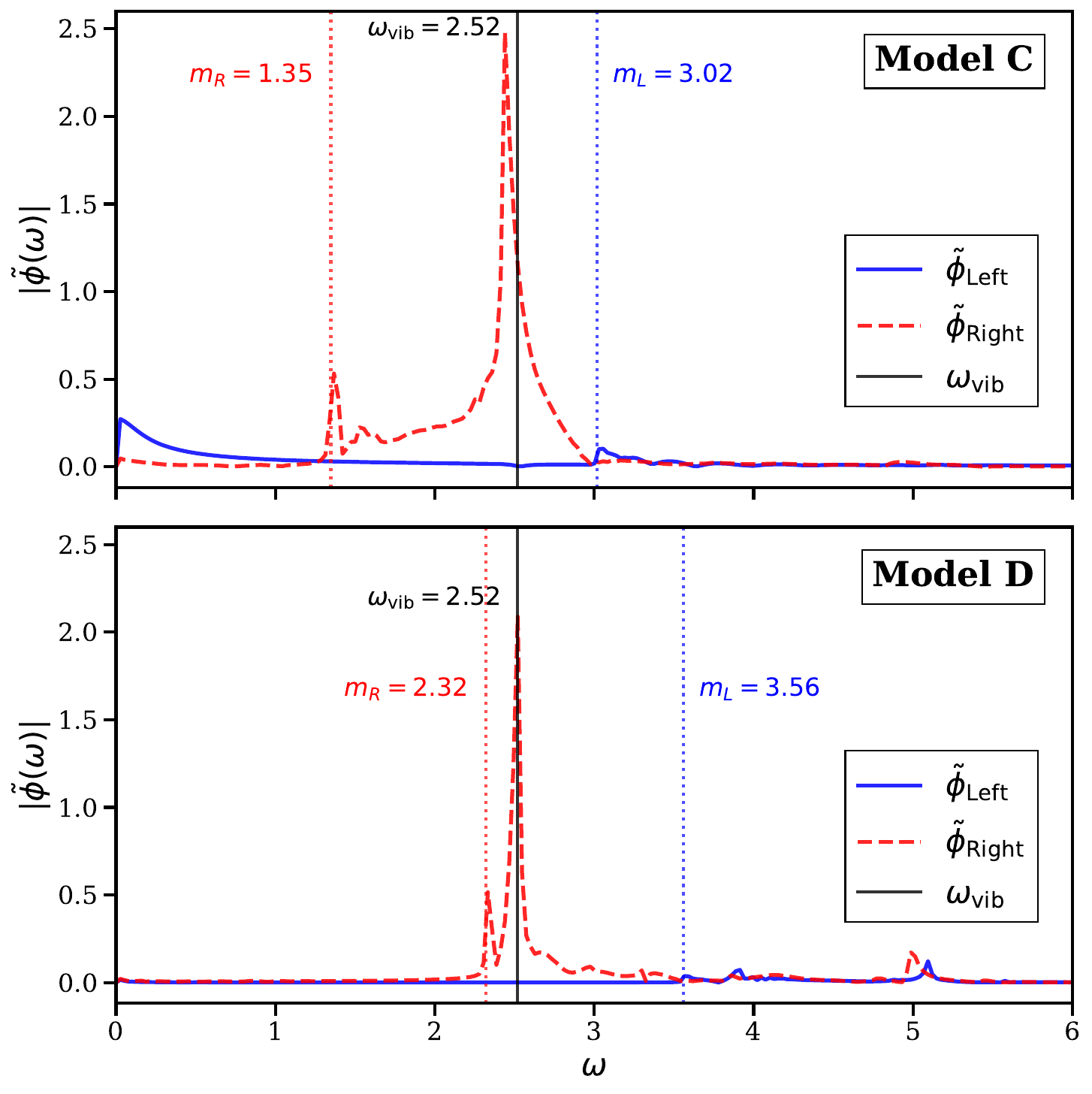}
    \caption{Same as Fig.~\ref{fig:fft_spectra_2D_AB}, but for the mass-asymmetric models (C and D). In these models, most of the power lies in the range $m_{\rm R} < \omega < m_{\rm L}$, suppressing propagation into the left vacuum and yielding predominantly rightward emission.}
    \label{fig:fft_spectra_2D_CD}
\end{figure}

\begin{figure}[htbp]
    \centering
    \includegraphics[width=0.85\textwidth]{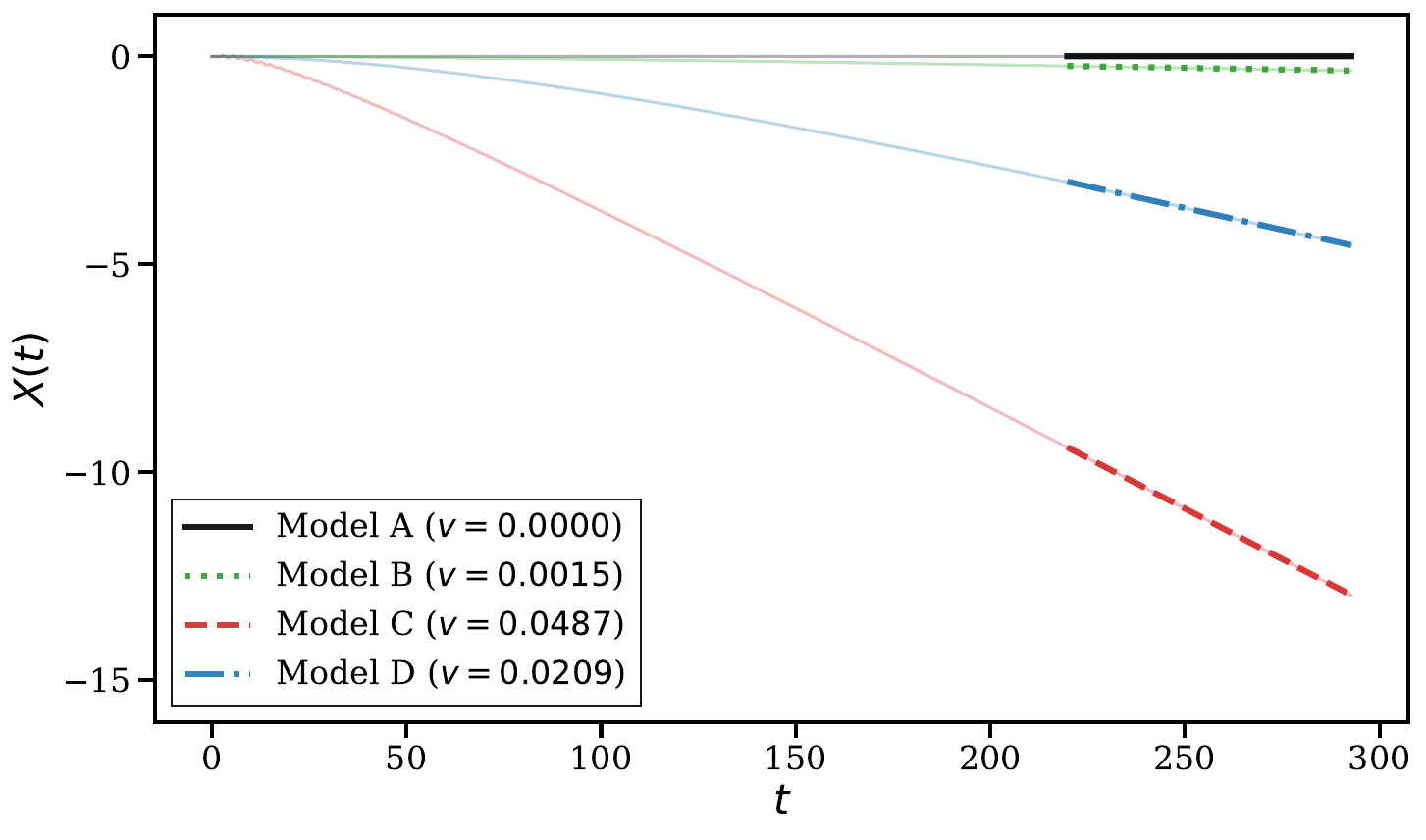}
    \caption{The mean wall position, $X(t)$, for the domain walls. The corresponding terminal velocities (shown in the legend) indicate that domain walls in the models with a vacuum mass differential (models C and D) develop a significant velocity drift and displacement growth. In contrast, domain walls in the skewed but mass-symmetric model B exhibit a much smaller velocity drift and displacement growth, while those in the baseline model A remain stationary throughout the evolution.}
    \label{fig:terminal_vel_2d}
\end{figure}

Unlike the $(1+1)$-dimensional case, where each wall vibrates at the intrinsic natural frequency of its corresponding shape mode, the extrinsic curvature associated with the wall wiggliness acts as a continuous driving force. 

The far-field scalar radiation spectra for models A and B are shown in Fig.~\ref{fig:fft_spectra_2D_AB}, together with the extracted internal vibrational frequency (black vertical line). In both models, the emitted radiation is sharply peaked at the driven vibrational frequency $\omega_{\rm vib}=2.52$. We note a very slight offset between the radiation peak and the internal vibrational frequency, which arises from the non-linear nature of domain wall dynamics and radiation emission. Since this frequency exceeds the scalar-field mass thresholds ($m_{\rm L}=m_{\rm R}=2$), the radiation propagates freely into both vacua. In model A, the emission is symmetric, whereas in model B it is only slightly asymmetric owing to the asymmetry of the potential barrier, which produces a small difference between the left- and right-going radiation fluxes.

For the mass-asymmetric models (C and D), the driven vibrational frequency lies between the two vacuum thresholds ($m_{\rm R}<\omega_{\rm vib}<m_{\rm L}$). The corresponding far-field scalar radiation spectra are shown in Fig.~\ref{fig:fft_spectra_2D_CD}. For these models, the observed rightwards radiation peaks are modified by a Doppler shift induced by the velocity of the walls. Because model C acquires a larger terminal velocity than model D, it exhibits a more pronounced Doppler shift to the left of $\omega_{\rm vib}$. As predicted by the $(1+1)$-dimensional analysis, scalar radiation is kinematically forbidden from propagating into the left vacuum and therefore flows almost exclusively into the right vacuum.

\subsection{Wall motion and recoil velocity}

At each time step, the domain wall position is identified as the $\phi=0$ level set of the scalar field. Specifically, the wall profile is described by the function $X(t,y)$ defined implicitly through
\begin{equation}
\phi[t,X(t,y),y]=0\,.
\end{equation}
The mean wall position, which we use to characterize the wall trajectory, is obtained by averaging over the transverse direction,
\begin{equation}
X(t)=\frac{1}{L}\int_0^L X(t,y)\,dy.
\end{equation}

The corresponding wall velocity, $v(t)$, is then obtained from a smoothed finite-difference derivative of $X(t)$. As a consistency check, the terminal recoil velocity was independently computed from global momentum conservation using Eq.~\eqref{velocity_theoretical}, following the $(1+1)$-dimensional construction generalized to integrations over the $(x,y)$ plane. The two determinations agree within $0.7\%$.

\begin{figure}[htbp]
    \centering
    \includegraphics[width=0.85\textwidth]{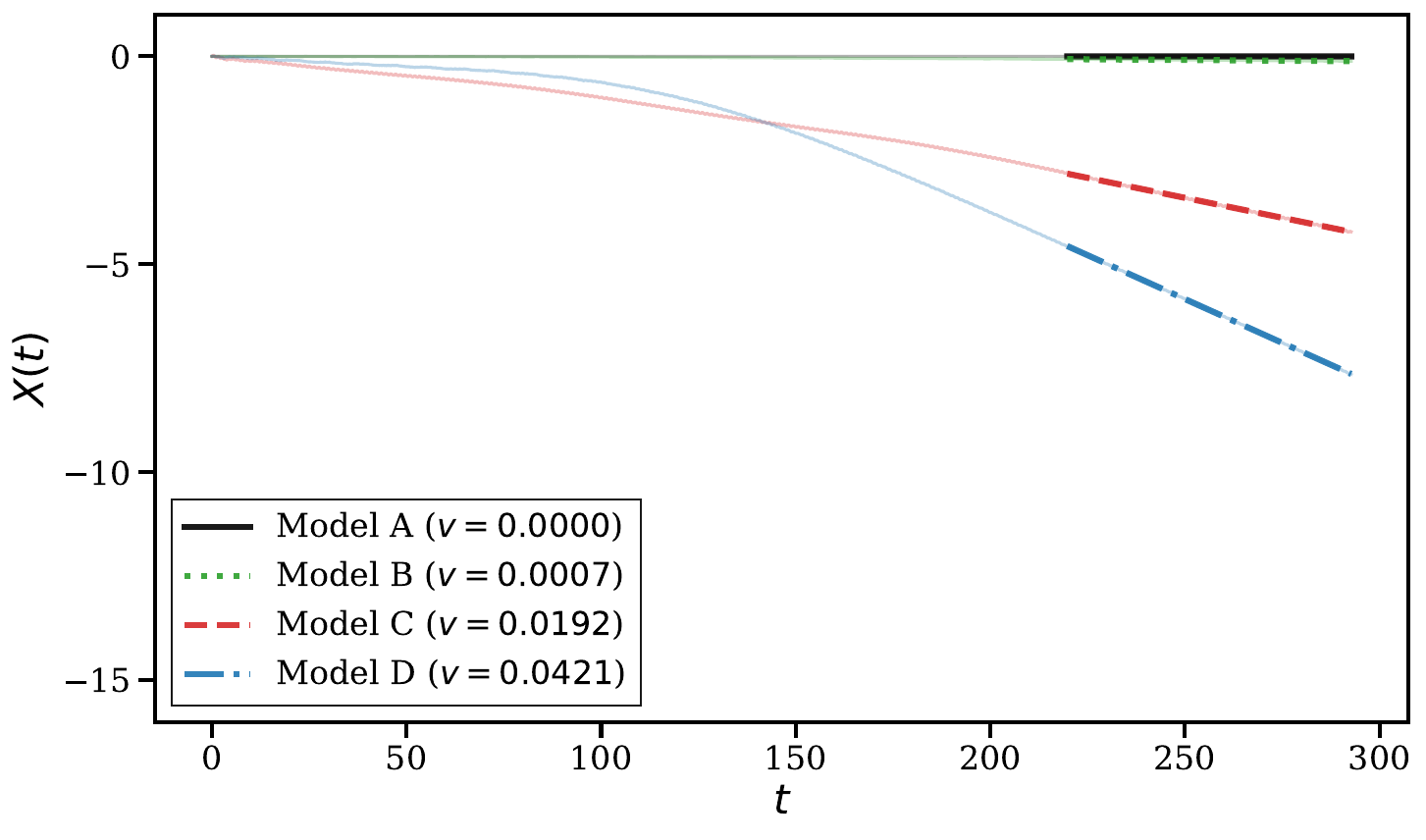}
    \caption{Similar to Fig.~\ref{fig:terminal_vel_2d}, but for the $\omega_{\rm vib}=5$ case. The qualitative behaviour is unchanged, although the recoil is slightly weaker.}
    \label{fig:terminal_vel_2d_5}
\end{figure}

The resulting averaged domain wall trajectories and the corresponding velocity fits are shown in Fig.~\ref{fig:terminal_vel_2d}. Model A remains stationary throughout the evolution, consistently with symmetric radiation emission. Model B exhibits only a small residual drift, with $v=-0.0015$. In contrast, the asymmetric radiation emission in models C and D induces a significantly larger recoil, driving the walls to terminal velocities of $v=-0.0487$ and $v=-0.0209$, respectively. This confirms that the vacuum-dependent scalar field mass is the primary cause of the rocket effect.

As in the $(1+1)$-dimensional case, the final kinetic energy of the domain wall is only a small fraction of the initially injected energy. For the more extreme mass-asymmetric model C, this fraction is approximately $2\%$. The recoil is again driven by the anisotropy of the emitted radiation, rather than by the total energy initially injected into the wall.

While the rocket effect is maximized when the internal vibrational frequency is trapped between the two vacuum mass thresholds ($m_{\rm R} < \omega < m_{\rm L}$), it is relevant to determine whether the bias is still present for resonances over both thresholds. To test this, we repeated the above simulation, but considering $\omega_{\rm wiggle}=2.5$, so that $\omega_{\rm vib}=5$. This frequency is enough to overcome the mass thresholds of both vacua in all models, consequently the radiation is no longer forbidden from entering the side with higher scalar field mass. Nonetheless, a bias favoring the right vacuum is still present (see Fig.~\ref{fig:terminal_vel_2d_5}). Because the left vacuum requires a higher excitation frequency for scalar fluctuations, waves transmitted into that region carry less linear momentum than those propagating into the right vacuum. This confirms that although kinematic blockage produces the strongest recoil, the vacuum mass asymmetry ($\Delta m^2$) consistently favors radiation emission into the lower-mass vacuum, thereby guaranteeing a persistent recoil toward the higher-mass side, irrespective of the excitation frequency.

\section{Rocket Effect in FLRW}

In this section, we investigate whether the rocket effect arising from the backreaction of anisotropic scalar radiation emitted by domain walls, described in the previous sections, can bias the evolution of cosmological domain wall networks and ultimately drive their decay. To this end, we perform $(2+1)$-dimensional numerical simulations of domain wall network evolution in a spatially flat FLRW universe for Models A–D.

\subsection{Cosmological Simulation Setup}

A well-known numerical challenge in numerical simulations of cosmic domain wall evolution is that, while their physical thickness remains constant, their comoving thickness decreases proportionally to $a^{-1}$. As a result, domain walls eventually become unresolved on a fixed comoving lattice. To overcome this issue, we employ the Press–Ryden–Spergel (PRS) algorithm \cite{Press:1989yh,Sousa:2010zza}, which modifies the field equation to maintain a constant comoving domain wall thickness while ensuring momentum conservation and the correct scaling properties of the network:
\begin{equation}
\frac{\partial^2 \phi}{\partial \eta^2}
+ 3 \mathcal{H} \frac{\partial \phi}{\partial \eta}
- \nabla^2 \phi
= - \frac{\partial V}{\partial \phi}\,.
\end{equation}
In this formulation, the damping coefficient is given by
\begin{equation}
\Gamma(\eta) = 3 \mathcal{H}\,,
\end{equation}
which yields $\Gamma = 3/\eta$ during the radiation dominated era ($a \propto \eta$) and $\Gamma = 6/\eta$ during the matter dominated era ($a \propto \eta^2$).

The Hubble friction term is evaluated using a centered Crank-Nicolson average. Integrating the equations of motion via a staggered leapfrog scheme in comoving coordinates yields the semi-implicit velocity update:
$$\begin{aligned}
\dot{\phi}_{i,j}^{[n+1/2]} &= \frac{\dot{\phi}_{i,j}^{[n-1/2]} \left(1 - \frac{1}{2} \Gamma^{[n]} \Delta \eta\right) + \Delta \eta \left[ (\nabla^2 \phi)_{i,j}^{[n]} - \left.\frac{\partial V}{\partial \phi}\right|_{\phi_{i,j}^{[n]}} \right]}{1 + \frac{1}{2} \Gamma^{[n]} \Delta \eta}\,, \\
\eta^{[n]} &= 1 + n\Delta\eta\,.
\end{aligned}$$
Here, the discrete time step index is $n \ge 0$, with the initial field configuration defined at $n=0$.

Simulations are performed on a two-dimensional spatial lattice with $2048^2$ grid points. The comoving box size is set to $L = 100$, corresponding to a spatial resolution of $\Delta x = 0.048$. The time evolution is carried out using a fixed time step $\Delta \eta = 0.2\,\Delta x$, ensuring stability of the staggered Leapfrog integration scheme while adequately resolving the domain wall dynamics. The system is evolved up to a final conformal time $\eta = 100$. Periodic boundary conditions are imposed in both spatial directions of the two-dimensional lattice.

\begin{figure}[htbp]
    \centering
    \includegraphics[width=0.85\textwidth]{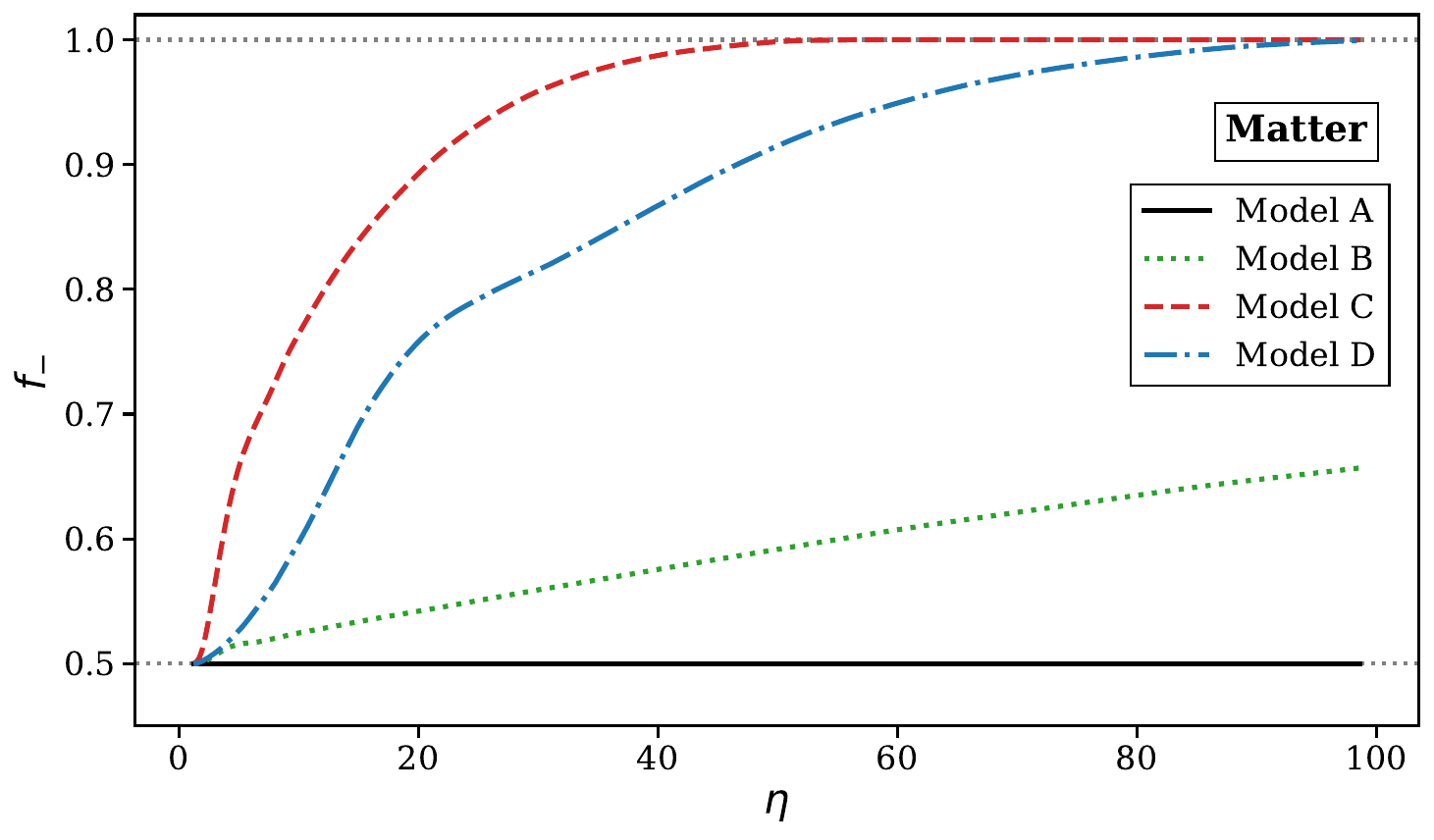}
    
    \vspace{0.4cm} 
    
    \includegraphics[width=0.85\textwidth]{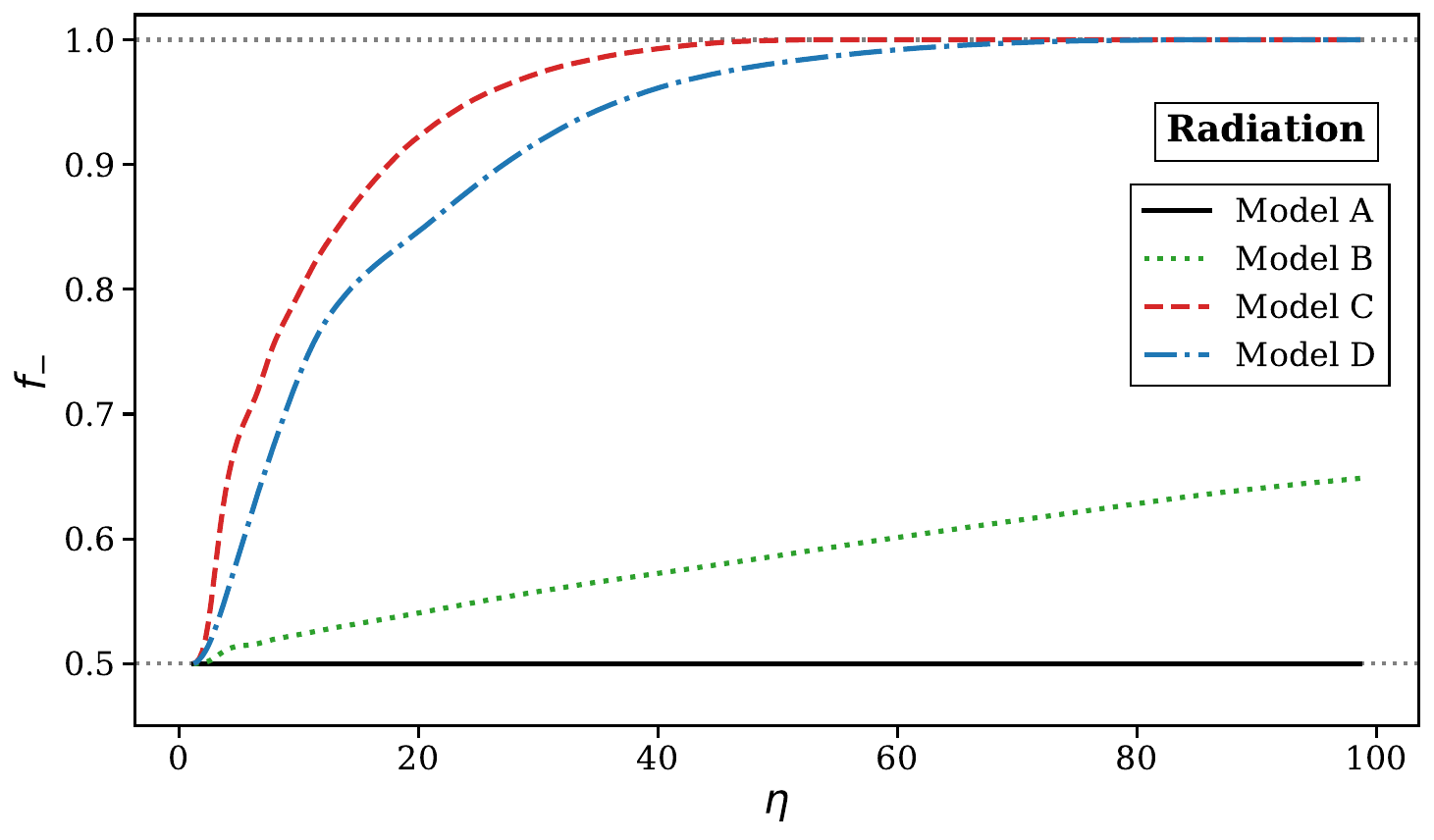}
    
    \caption{Evolution of the volume fraction of the vacuum with smaller scalar field mass ($f_-$) as a function of conformal time $\eta$ in a matter-dominated (top panel) and radiation-dominated (bottom panel) FLRW universe. In both expansion regimes, models with $\Delta m^2 \neq 0$ (C and D) systematically favor the vacuum with smaller scalar field mass. The symmetric reference model (A) remains at $f_- = 0.5$ throughout the evolution, confirming the absence of spurious numerical bias, while the skew-only model with $\Delta m^2 = 0$ (B) develops a significantly weaker, but clearly identifiable, volumetric bias.}
    \label{fig:cosmo_fraction_combined}
\end{figure}

Initial conditions are generated from a Gaussian white-noise field with a mean of $\mu = 0.0$ and variance of $\sigma^2 = 1.0$. We smooth this raw noise using a spatial Gaussian filter with a standard deviation of $\sigma_s = 3.0 \Delta x$. The resulting continuous field is then projected onto the two degenerate vacuum states by thresholding at the median value: lattice sites with field values above (below) the median are assigned to one (the other) vacuum. This procedure yields a statistically balanced initial configuration with a finite correlation length, such that each vacuum occupies a volume fraction of $1/2$ on average.

To reduce statistical biases associated with individual realizations of the initial conditions, we employ an antithetic pairing strategy. For each of 13 random seeds, two simulations with opposite vacuum assignments are performed, and the observable is obtained by averaging over each pair. This procedure removes the asymmetry associated with the initial choice of vacua, mitigating the impact of initial-condition biases and isolating the physical effects driven by the underlying model dynamics.

\begin{table}[htbp]
    \centering
    \caption{Summary of the derived quantities for all models considered in the cosmological domain wall network simulations. Again, the normalization parameter $\lambda$ is tuned to enforce strict domain wall tension degeneracy ($\sigma = 4/3$) across all models.} 
    \label{tab:cosmo_parameters}
    \renewcommand{\arraystretch}{1.3}
    \begin{tabular}{lcccc}
    \toprule\toprule
    \textbf{Model} & $m_{\rm L}$ & $m_{\rm R}$ & $\Delta m^2$ & $d3V$ \\
    \midrule
    \ \ \ \ A   & $2.000$ & $2.000$ & $0.000$ & $0.000$ \\
    \ \ \ \ B   & $1.999$ & $1.999$ & $0.000$ & $0.100$ \\
    \ \ \ \ C   & $2.033$ & $1.967$ & $0.267$ & $0.100$ \\
    \ \ \ \ D   & $2.981$ & $2.936$ & $0.267$ & $0.000$ \\
    \bottomrule\bottomrule
    \end{tabular}
\end{table}

To avoid immediate decay of the network, cosmological simulations employ less asymmetric potentials than those used in the Minkowski case (see Table~\ref{tab:cosmo_parameters}), ensuring that the network persists long enough for its dynamical evolution to be properly resolved.

\subsection{Biased Cosmological Domain Wall Dynamics}

We use the volume fraction of the vacuum with smaller scalar field mass, denoted by $f_-$, as a measure of the bias in the cosmological evolution of domain walls with asymmetric potentials. All simulations are initialized with $f_-=0.5$. 

Figure~\ref{fig:cosmo_fraction_combined} shows the evolution of $f_-$ as a function of conformal time in both radiation- and matter-dominated backgrounds. A clear and systematic trend is observed: models with asymmetric excitation spectra develop a progressive imbalance, with the lower-mass vacuum occupying an increasingly larger fraction of the volume. This behaviour is qualitatively identical in both cosmological eras. 

The symmetric reference model (Model A) remains stable at $f_-=0.5$, confirming that the numerical setup does not introduce spurious biases. In contrast, models with a mass asymmetry (Models C and D) exhibit a strong, monotonic drift toward the lower-mass vacuum. The skew-only model (Model B) also develops a volumetric bias, albeit at a significantly slower rate, demonstrating that the vacuum dependence of the scalar field mass, rather than the asymmetry of the potential barrier near its local maximum, is the primary driver of the biased evolution. We have also verified that varying the initial correlation length by increasing (decreasing) the width $\sigma_{\rm s}$ of the Gaussian filter leads to a slower (faster) evolution of the volume fraction $f_-$ without affecting either the direction of the bias or our overall conclusions. 

These results demonstrate the cosmological impact of the rocket effect associated with asymmetric scalar radiation emission. The resulting recoil force biases the evolution of cosmological domain wall networks in favor of the lower-mass vacuum, thereby promoting the decay of the network. In theories with degenerate vacua, this mechanism alone is sufficient to generate a dynamical bias, even in the absence of any differences in vacuum energy densities. 

More generally, in theories with non-degenerate vacua, the rocket effect constitutes a complementary source of dynamical bias that acts alongside the conventional volume pressure arising from vacuum energy differences. Depending on whether the recoil force acts in the same or the opposite direction as the conventional volume pressure, it can either hasten or delay network decay relative to the standard expectation based solely on vacuum energy differences. Interestingly, recent high-resolution field-theory simulations~\cite{Babichev:2025stm,Cyr:2025nzf}, performed in models with both non-degenerate vacuum energy densities and vacuum-dependent scalar field masses, have shown that the network decay time exhibits a non-trivial dependence on the vacuum energy density difference $\delta V$, differing from the naive estimate $t_{\rm decay}\sim\sigma/\delta V$. Our results suggest that this behaviour cannot, in general, be attributed solely to the vacuum energy difference, since the recoil mechanism associated with the vacuum dependence of the scalar field mass provides an additional contribution to the dynamical bias that can modify the dependence of the decay time on $\delta V$.

Recently, an effective-action description of domain-wall dynamics including higher-order curvature corrections to the Nambu-Goto action was developed in \cite{Blanco-Pillado:2024bev}. These corrections can modify the dynamics of curved walls and, for asymmetric wall profiles, give rise to an acceleration whose direction changes when the two vacua on either side of the wall are interchanged. Although this effect is distinct from the rocket effect considered here, which is already present for planar walls with vanishing extrinsic curvature, both effects may contribute simultaneously to the dynamics of curved walls with asymmetric profiles.

We also note that the different spectra of linearized fluctuations around the two vacua might generate an additional quantum bias, even when the vacua are energetically degenerate at the classical level \cite{Weigel:2017iup, Takyi:2020tvl}. This effect is distinct from the classical rocket effect considered here: the latter is a classical effect resulting from the asymmetric emission of scalar radiation and the associated momentum flux, whereas the former originates from vacuum fluctuations and is associated with a quantum correction to the classical dynamics.

\section{Conclusions}\label{sec:conc}

We have investigated the dynamics of domain walls in scalar field theories with vacuum-dependent scalar field masses. Combining analytical arguments with numerical simulations in both $1+1$ and $2+1$ dimensions, we have shown that asymmetries in the vacuum excitation spectrum lead to anisotropic emission of scalar radiation, preferentially directed toward the vacuum with the smaller scalar field mass. This asymmetry arises from vacuum-dependent kinematic constraints associated with a non-vanishing mass splitting and does not require any difference in the vacuum energy densities. The resulting asymmetric momentum flux generates a recoil (rocket) force on the domain wall, driving it toward the vacuum with the larger scalar field mass.

Our results further demonstrate that this recoil mechanism, acting alone, can significantly influence the large-scale evolution of cosmological domain wall networks. By preferentially shrinking domains associated with the vacuum of larger scalar field mass, it may drive network decay even in models with degenerate vacua, thereby providing an additional dynamical mechanism for resolving the cosmological domain wall problem. More generally, in theories with non-degenerate vacua, the rocket effect constitutes a complementary source of dynamical bias that acts alongside the pressure generated by vacuum energy differences. Depending on its direction relative to the conventional volume pressure, it can either hasten or delay network decay compared with the standard expectation based solely on vacuum energy differences.

An important implication of our analysis is the reinterpretation of previously reported biased domain wall evolution in models with asymmetric potentials. We find that the observed bias, previously attributed to asymmetries in the potential barrier near the local maximum, is instead primarily controlled by the vacuum dependence of the scalar field mass. This result clarifies the physical origin of the bias in such models and establishes the rocket effect as a generic and potentially significant ingredient in the dynamics of cosmological domain wall networks whenever the scalar excitation spectrum differs between the vacua.

\acknowledgments

We thank Clara Winckler and Lara Sousa for many valuable discussions and for drawing our attention to Refs.~\cite{Babichev:2025stm,Cyr:2025nzf}, whose high-resolution field-theory simulations reveal a non-trivial dependence of the domain wall network decay time on the vacuum energy density difference.
We also thank Adalto Gomes and our colleagues in the Cosmology Group at the Instituto de Astrofísica e Ciências do Espaço for insightful discussions. This work was supported by the Fundação para a Ciência e a Tecnologia (FCT) through the research grants UID/04434/2025 (DOI: \href{https://doi.org/10.54499/UID/04434/2025}{10.54499/UID/04434/2025}) and 2024.17828.PEX -- \emph{Unveiling the Early Universe with Topological Defects} (DOI: \href{https://doi.org/10.54499/2024.17828.PEX}{10.54499/2024.17828.PEX}).
	
\bibliography{Rocket}

\begin{thebibliography}{10}

\bibitem{NANOGrav:2023hvm}
Adeela Afzal et~al.
\newblock {The NANOGrav 15 yr Data Set: Search for Signals from New Physics}.
\newblock {\em Astrophys. J. Lett.}, 951(1):L11, 2023.
\newblock [Erratum: Astrophys.J.Lett. 971, L27 (2024), Erratum: Astrophys.J. 971, L27 (2024)].

\bibitem{Avelino:2008mh}
P.~P. Avelino, D.~Bazeia, R.~Menezes, and J.~Oliveira.
\newblock {Bifurcation and pattern changing with two real scalar fields}.
\newblock {\em Phys. Rev. D}, 79:085007, 2009.

\bibitem{Avelino:2008qy}
P.~P. Avelino, C.~J. A.~P. Martins, and L.~Sousa.
\newblock {Dynamics of Biased Domain Walls and the Devaluation Mechanism}.
\newblock {\em Phys. Rev. D}, 78:043521, 2008.

\bibitem{Babichev:2025stm}
E.~Babichev, I.~Dankovsky, D.~Gorbunov, S.~Ramazanov, and A.~Vikman.
\newblock {Biased domain walls: faster annihilation, weaker gravitational waves}.
\newblock {\em JCAP}, 10:103, 2025.

\bibitem{Blanco-Pillado:2024bev}
Jose~J. Blanco-Pillado, Alberto Garc{\'\i}a Mart{\'\i}n-Caro, Daniel Jim{\'e}nez-Aguilar, and Jose~M. Queiruga.
\newblock {Effective actions for domain wall dynamics}.
\newblock {\em Phys. Rev. D}, 111(5):056007, 2025.

\bibitem{Blanco-Pillado:2022rad}
Jose~J. Blanco-Pillado, Daniel Jim{\'e}nez-Aguilar, Jose~M. Queiruga, and Jon Urrestilla.
\newblock {The dynamics of domain wall strings}.
\newblock {\em JCAP}, 05:011, 2023.

\bibitem{Blasi:2022ayo}
Simone Blasi, Alberto Mariotti, A{\"a}ron Rase, Alexander Sevrin, and Kevin Turbang.
\newblock {Friction on ALP domain walls and gravitational waves}.
\newblock {\em JCAP}, 04:008, 2023.

\bibitem{Bucher:1998mh}
Martin Bucher and David~N. Spergel.
\newblock {Is the dark matter a solid?}
\newblock {\em Phys. Rev. D}, 60:043505, 1999.

\bibitem{Caloni:2026dyu}
Luca Caloni, Ricardo~Z. Ferreira, Lara Sousa, and Clara Winckler.
\newblock {Cosmic strings and domain walls: the impact of CMB $B$-mode data}.
\newblock 2 2026.

\bibitem{Correia:2014kqa}
J.~R. C. C.~C. Correia, I.~S. C.~R. Leite, and C.~J. A.~P. Martins.
\newblock {Effects of Biases in Domain Wall Network Evolution}.
\newblock {\em Phys. Rev. D}, 90(2):023521, 2014.

\bibitem{Correia:2018tty}
J.~R. C. C.~C. Correia, I.~S. C.~R. Leite, and C.~J. A.~P. Martins.
\newblock {Effects of biases in domain wall network evolution. II. Quantitative analysis}.
\newblock {\em Phys. Rev. D}, 97(8):083521, 2018.

\bibitem{Coulson:1995nv}
D.~Coulson, Z.~Lalak, and Burt~A. Ovrut.
\newblock {Biased domain walls}.
\newblock {\em Phys. Rev. D}, 53:4237--4246, 1996.

\bibitem{Cyr:2025nzf}
Bryce Cyr, Steven Cotterill, and Richard Battye.
\newblock {Near-peak spectrum of gravitational waves from collapsing domain walls}.
\newblock {\em Phys. Rev. D}, 113(4):043549, 2026.

\bibitem{Dunsky:2024zdo}
David~I. Dunsky and Marius Kongsore.
\newblock {Primordial black holes from axion domain wall collapse}.
\newblock {\em JHEP}, 06:198, 2024.

\bibitem{Ferreira:2024eru}
Ricardo~Z. Ferreira, Alessio Notari, Oriol Pujol\`as, and Fabrizio Rompineve.
\newblock {Collapsing domain wall networks: impact on pulsar timing arrays and primordial black holes}.
\newblock {\em JCAP}, 06:020, 2024.

\bibitem{Filippov:2025dpb}
D.~P. Filippov and A.~A. Kirillov.
\newblock {Interaction of Domain Walls with Scalar Particles in the Early Universe}.
\newblock {\em Phys. Atom. Nucl.}, 88(3):540--545, 2025.

\bibitem{Forgacs:2008az}
Peter Forgacs, Arpad Lukacs, and Tomasz Romanczukiewicz.
\newblock {Negative radiation pressure exerted on kinks}.
\newblock {\em Phys. Rev. D}, 77:125012, 2008.

\bibitem{Gouttenoire:2023ftk}
Yann Gouttenoire and Edoardo Vitagliano.
\newblock {Domain wall interpretation of the PTA signal confronting black hole overproduction}.
\newblock {\em Phys. Rev. D}, 110(6):L061306, 2024.

\bibitem{Gruber:2024pqh}
D.~Gr\"uber, L.~Sousa, and P.~P. Avelino.
\newblock {Stochastic gravitational wave background generated by domain wall networks}.
\newblock {\em Phys. Rev. D}, 110(2):023505, 2024.

\bibitem{Hindmarsh:1996xv}
Mark Hindmarsh.
\newblock {Analytic scaling solutions for cosmic domain walls}.
\newblock {\em Phys. Rev. Lett.}, 77:4495--4498, 1996.

\bibitem{Hiramatsu:2013qaa}
Takashi Hiramatsu, Masahiro Kawasaki, and Ken'ichi Saikawa.
\newblock {On the estimation of gravitational wave spectrum from cosmic domain walls}.
\newblock {\em JCAP}, 02:031, 2014.

\bibitem{Kibble:1976sj}
T.~W.~B. Kibble.
\newblock {Topology of Cosmic Domains and Strings}.
\newblock {\em J. Phys. A}, 9:1387--1398, 1976.

\bibitem{Kitajima:2023kzu}
Naoya Kitajima, Junseok Lee, Fuminobu Takahashi, and Wen Yin.
\newblock {Stability of domain walls with inflationary fluctuations under potential bias, and gravitational wave signatures}.
\newblock {\em JCAP}, 07:053, 2025.

\bibitem{Krajewski:2021jje}
Tomasz Krajewski, Jan~Henryk Kwapisz, Zygmunt Lalak, and Marek Lewicki.
\newblock {Stability of domain walls in models with asymmetric potentials}.
\newblock {\em Phys. Rev. D}, 104(12):123522, 2021.

\bibitem{Larsson:1996sp}
Sebastian~E. Larsson, Subir Sarkar, and Peter~L. White.
\newblock {Evading the cosmological domain wall problem}.
\newblock {\em Phys. Rev. D}, 55:5129--5135, 1997.

\bibitem{Li:2025gld}
Yuan-Jie Li, Jing Liu, and Zong-Kuan Guo.
\newblock {Dynamics of ZN domain walls with bias directions}.
\newblock {\em Phys. Rev. D}, 112(10):103510, 2025.

\bibitem{newman2013computational}
Mark E.~J. Newman.
\newblock {\em Computational Physics}.
\newblock CreateSpace Independent Publishing Platform, 2013.

\bibitem{Notari:2025kqq}
Alessio Notari, Fabrizio Rompineve, and Francisco Torrenti.
\newblock The spectrum of gravitational waves from annihilating domain walls.
\newblock {\em JCAP}, 07:049, 2025.

\bibitem{PinaAvelino:2006ia}
Pedro Pina~Avelino, C.~J. A.~P. Martins, J.~Menezes, R.~Menezes, and J.~C. R.~E. Oliveira.
\newblock {Frustrated expectations: defect networks and dark energy}.
\newblock {\em Phys. Rev. D}, 73:123519, 2006.

\bibitem{Press:1989yh}
William~H. Press, Barbara~S. Ryden, and David~N. Spergel.
\newblock {Dynamical Evolution of Domain Walls in an Expanding Universe}.
\newblock {\em Astrophys. J.}, 347:590--604, 1989.

\bibitem{press2007numerical}
William~H. Press, Saul~A. Teukolsky, William~T. Vetterling, and Brian~P. Flannery.
\newblock {\em Numerical Recipes 3rd Edition: The Art of Scientific Computing}.
\newblock Cambridge University Press, 3 edition, 2007.

\bibitem{Romanczukiewicz:2017hdu}
Tomasz Roma{\'n}czukiewicz.
\newblock {Could the primordial radiation be responsible for vanishing of topological defects?}
\newblock {\em Phys. Lett. B}, 773:295--299, 2017.

\bibitem{Rubin:2001yw}
Sergei~G. Rubin, Alexander~S. Sakharov, and Maxim~Yu. Khlopov.
\newblock {The Formation of primary galactic nuclei during phase transitions in the early universe}.
\newblock {\em J. Exp. Theor. Phys.}, 91:921--929, 2001.

\bibitem{saad2011numerical}
Yousef Saad.
\newblock {\em Numerical Methods for Large Eigenvalue Problems}.
\newblock Society for Industrial and Applied Mathematics, Philadelphia, PA, 2011.

\bibitem{Sousa:2010zza}
L.~Sousa and P.~P. Avelino.
\newblock {Evolution of domain wall networks: The Press-Ryden-Spergel algorithm}.
\newblock {\em Phys. Rev. D}, 81:087305, 2010.

\bibitem{Sousa:2015cqa}
L.~Sousa and P.~P. Avelino.
\newblock {Cosmic Microwave Background anisotropies generated by domain wall networks}.
\newblock {\em Phys. Rev. D}, 92(8):083520, 2015.

\bibitem{Takyi:2020tvl}
I.~Takyi, M.~K. Matfunjwa, and H.~Weigel.
\newblock {Quantum corrections to solitons in the $\Phi^8$ model}.
\newblock {\em Phys. Rev. D}, 102(11):116004, 2020.

\bibitem{thomas1995numerical}
J.~W. Thomas.
\newblock {\em Numerical Partial Differential Equations: Finite Difference Methods}, volume~22 of {\em Texts in Applied Mathematics}.
\newblock Springer New York, 1995.

\bibitem{Vilenkin:2000jqa}
A.~Vilenkin and E.~P.~S. Shellard.
\newblock {\em {Cosmic Strings and Other Topological Defects}}.
\newblock Cambridge University Press, 7 2000.

\bibitem{Vilenkin:1984ib}
Alexander Vilenkin.
\newblock {Cosmic Strings and Domain Walls}.
\newblock {\em Phys. Rept.}, 121:263--315, 1985.

\bibitem{Weigel:2017iup}
H.~Weigel.
\newblock {Emerging Translational Variance: Vacuum Polarization Energy of the $\phi^6$ Kink}.
\newblock {\em Adv. High Energy Phys.}, 2017:1486912, 2017.

\bibitem{Winckler:2025hbc}
Clara Winckler, Pedro~P. Avelino, and Lara Sousa.
\newblock {Biased domain walls and the origin of early massive structures}.
\newblock {\em Phys. Rev. D}, 112(12):123506, 2025.

\bibitem{Zeldovich:1974uw}
Ya.~B. Zeldovich, I.~Yu. Kobzarev, and L.~B. Okun.
\newblock {Cosmological Consequences of the Spontaneous Breakdown of Discrete Symmetry}.
\newblock {\em Zh. Eksp. Teor. Fiz.}, 67:3--11, 1974.

\end{thebibliography}
	
\end{document}